# Symbolic Dynamics and Periodic Orbits for the Cardioid Billiard

by

A. Bäcker[†1)] and H. R. Dullin[‡ 2)]

1) II. Institut für Theoretische Physik, Universität Hamburg
Luruper Chaussee 149, D–22761 Hamburg
Federal Republic of Germany

2) Institut für Dynamische Systeme and Institut für Theoretische Physik,
NW 1, Universität Bremen
Postfach 330 440, D–28334 Bremen
Federal Republic of Germany

**Abstract:**

The periodic orbits of the strongly chaotic cardioid billiard are studied by introducing a binary symbolic dynamics. The corresponding partition is mapped to a topological well-ordered symbol plane. In the symbol plane the pruning front is obtained from orbits running either into or through the cusp. We show that all periodic orbits correspond to maxima of the Lagrangian and give a complete list up to code length 15. The symmetry reduction is done on the level of the symbol sequences and the periodic orbits are classified using symmetry lines. We show that there exists an infinite number of families of periodic orbits accumulating in length and that all other families of geometrically short periodic orbits eventually get pruned. All these orbits are related to finite orbits starting and ending in the cusp. We obtain an analytical estimate of the Kolmogorov-Sinai entropy and find good agreement with the numerically calculated value and the one obtained by averaging periodic orbits. Furthermore the statistical properties of periodic orbits are investigated.

[†]Supported by Deutsche Forschungsgemeinschaft under Contract No. DFG-Ste 241/7-1
On leave of absence from: Universität Ulm, Abteilung für Theoretische Physik, Albert-Einstein-Allee 11, 89069 Ulm, E-mail address: `baec@physik.uni-ulm.de`
[‡]Supported by Deutsche Forschungsgemeinschaft, E-mail address: `hdullin@physik.uni-bremen.de`

# 1 Introduction

A key step towards an understanding of the behaviour of a dynamical system is achieved by finding a symbolic dynamics. By means of the symbolic dynamics trajectories can be labeled by doubly infinite sequences built from a (not necessarily) finite alphabet. Periodic orbits are represented by periodic sequences and can be systematically searched for once the coding is known. The knowledge of a complete set of a large number of periodic orbits up to a given geometric length is necessary for the application of Gutzwillers periodic orbit theory [1], which relates the quantum mechanical density of states of the quantized billiard system to a sum over classical periodic orbits. If the symbolic description is not complete there are sequences which do not correspond to physical orbits. To deal with this more complicated case the idea of a pruning front in the symbol plane was introduced in [2]. These methods have been successfully applied to a number of systems. There are, however, no systematic methods to find a symbolic description and the corresponding pruning mechanisms.

In the class of billiards inside simply connected domains of the Euclidean plane ergodic examples typically have either families of orbits accumulating in length, singularities in the boundary or non-isolated parabolic families. In the case of the cardioid billiard, which has been rigorously proven to be strongly chaotic, i.e. it is ergodic, mixing, a $K$-system and even a Bernoulli system [3, 4, 5, 6], we have accumulating families and one singularity. The relation of the two in the cardioid billiard is quite interesting and a thorough understanding of their effects is a prerequisite for the semiclassical quantization of this system.

There already are several results for both, the classical and quantum mechanical billiard. The cardioid is the limiting case of a family of billiards introduced by Robnik [7], see also [8, 9] and references therein. The statistical properties of the eigenvalues of the quantized cardioid billiard were studied in detail in [10, 11]; see also [12], where the focus is on diffraction effects. A lot of work has been done on Robnik's family but the classical mechanics of the cardioid has not been analyzed in depth. It is this gap we want to fill in with this work, keeping in mind, however, the application to quantum mechanics.

The paper is organized as follows. In sec. 2 the cardioid billiard and the billiard map are defined. We show that products of linearized maps always have a positive trace. Therefore all orbits are maxima of the Lagrangian. Subsequently a discussion of the symmetries of the billiard map is given. In sec. 3 the symbolic dynamics is defined and the corresponding partition of the Poincaré section is illustrated. The initial partition is given by the discontinuity of the map. The pruning of code words is discussed in the symbol plane using the pruning front, which is also obtained from the discontinuity of the map. In sec. 4 the periodic orbits are classified according to their symmetry by using the symmetry lines of the desymmetrized billiard. We give a list of the number of periodic orbits in each symmetry class up to code length 15. Families of periodic orbits with short geometric length and their relation to cusp orbits are investigated next. It is shown that most of them eventually get pruned. However, there remains an infinite number of families accumulating in length. Concerning the number of cusp orbits we show that they are more frequent than periodic orbits. In sec. 5 we obtain an analytical estimate for the Kolmogorov-Sinai entropy and find good agreement with the value obtained from numerical methods. The average length and the KS entropy are calculated using the periodic orbits. Finally, we focus on the statistics of periodic orbits.

# 2 The cardioid billiard

A billiard inside a two dimensional Euclidean domain $\Omega$ is given by the free motion of a point particle inside $\Omega$ with elastic reflections at the boundary $\partial\Omega$, i.e. the angle of incidence equals the angle of reflection. The cardioid billiard is the limiting case of a family of billiards first studied by Robnik [7]. Their boundary in polar coordinates $(\rho, \phi)$ is given by

$$\rho(\phi) = 1 + \epsilon \cos \phi, \qquad \phi \in [-\pi, \pi]. \tag{1}$$

We restrict our attention to the cardioid (see fig. 1) which is obtained for $\epsilon = 1$, or implicitly by

$$F(x, y) = (x^2 + y^2 - x)^2 - (x^2 + y^2) = 0, \tag{2}$$

where $(x, y) = r(\phi) = (\rho(\phi) \cos \phi, \rho(\phi) \sin \phi)$. At $\phi = \pm\pi$ the cardioid has a singularity located at the origin $r(\pm\pi) = (0, 0)$.

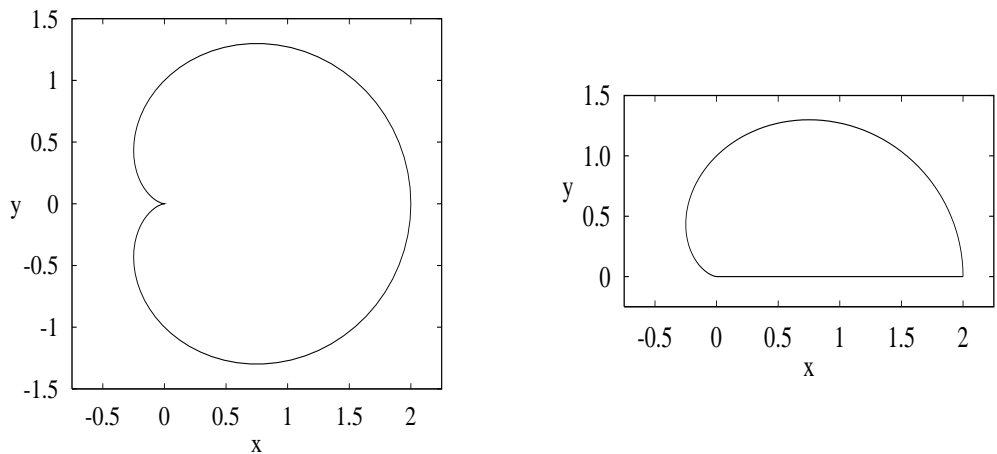

Figure 1: The full and the desymmetrized cardioid billiard.

From the above definition (1) one easily derives the curvature

$$\kappa(\phi) = \frac{3}{4 \cos(\phi/2)}, \tag{3}$$

the unit tangent vector

$$T = (T_x, T_y) = (-\sin(3\phi/2), \cos(3\phi/2)), \tag{4}$$

and the differential of the arc length $\frac{ds}{d\phi} = 2\cos(\phi/2)$. Thus the arc length $s$ is related to $\phi$ by

$$s = 4 \sin(\phi/2) \ . \tag{5}$$

The area of the cardioid is $|\Omega| = 3\pi/2$ and its circumference is $|\partial\Omega| = 8$.

## 2.1 Poincaré map

We now derive the Poincaré map from bounce to bounce in Birkhoff coordinates $\xi = (s, p)$ where $p$ is the component of the velocity in the direction of $T$ right after the reflection. The Cartesian components of the unit velocity $v$ of a billiard ball starting on $\partial \Omega$ at $r(\phi)$ is determined by the angle $\beta \in [pi/2, \pi/2]$ measured with respect to the normal $N = (-T_y, T_x)$ pointing inward; $\beta = -\pi/2, 0, \pi/2$ specifies the direction $-T, N, T$ respectively. The velocity in the $T, N$ coordinate system is denoted by $(p, n) = (\sin \beta, \cos \beta)$, so that we obtain $v = (-\cos(\beta - 3\phi/2), \sin(\beta - 3\phi/2))$. The rightmost point of $\partial \Omega$ corresponds to arc length $s = 0$, so that $s$ extends from $-4$ to $4$ while $p \in [-1, 1]$. We now construct the Poincaré map $P$ from the rectangle $\mathcal{P} = [-4, 4] \times [-1, 1]$ onto itself by calculating the image of a point $\xi = (s, p) \in \mathcal{P}$. We seek the first intersection of the ray starting at $r = (x, y)$ in direction $v$ with the boundary curve $\partial \Omega$

$$F(x + tv_x, y + tv_y) = 0, \quad F(x, y) = 0. \tag{6}$$

One root of this fourth order polynomial in $t$ is always zero, because of $F(x, y) = 0$. Thus we obtain a polynomial $Q(t)$ of degree 3 for t

$$\begin{aligned} Q(t) &= t^3 - \left(4\cos(-\beta + \frac{\phi}{2}) + 2\cos(\beta + \frac{\phi}{2})\right)t^2 \\ &\quad + (7 + 8\cos(\phi) + 4\cos(-2\beta + \phi) + \cos(2\beta + \phi) + 4\cos(2\beta))\,t/2 \\ &\quad - 8\cos(\beta)\cos^3(\frac{\phi}{2})\ . \end{aligned} \tag{7}$$

The smallest zero $t^\star > 0$ is the physically meaningful solution (only for the sliding motion we have $t^\star = 0$). To complete the description of the Poincaré map we first calculate the new point $r' = (x', y') = (x + t^\star v_x, y + t^\star v_y)$ on $\partial \Omega$ and determine

$$\begin{aligned} \phi' &= \arctan(y'/x') \\ p' &= \sin(\beta') = \langle T', v' \rangle = \langle T', v \rangle = \sin(3(\phi' - \phi)/2 + \beta)\ . \end{aligned} \tag{8}$$

The complete map $P : \xi = (s, p) \mapsto \xi' = (s', p')$ is now given by the composition of the above steps. $P$ is area preserving because $s$ and $p$ are canonically conjugate. For convenience we sometimes use $\phi$ instead of $s$, but without mentioning we assume it to be expressed in terms of $s$.

The two curves

$$\mathcal{S}_\pm = \{\xi \in \mathcal{P} \mid s = \pm 4\}, \tag{9}$$

corresponds to orbits which start in the singularity $\phi = \pm \pi$, i.e. on the right or left boundary of $\mathcal{P}$. For the image of points $\xi = (s, p) \in \mathcal{S}_\pm$ under $P$ one has $\phi' = \arctan(v_y/v_x)$ because of $r = (0, 0)$. Thus one has $\phi' = \pm \pi/2 - \beta$ and using equation (8) $p' = -\sin(\pm \pi/4 - \beta/2)$. The two image curves join at the origin of $\mathcal{P}$. For reasons that become clear later on, we denote them by

$$\Gamma_\pm^{-1} = \{\xi \in \mathcal{P} \mid p = -s/4,\ \pm s \geq 0\}\ , \tag{10}$$

such that

$$P(\mathcal{S}_\pm) = \Gamma_\pm^{-1}. \tag{11}$$

Each of the curves $\mathcal{S}_+$ and $\mathcal{S}_-$ has a fixed point $(4, -1)$ respectively $(-4, 1)$, which corresponds to $v_x = -1$. The physical motion starting at these fixed points is a sliding motion along the boundary, either counterclockwise or clockwise.

Note that although $r(-\pi) = r(\pi) = (0, 0)$ we take them as different points equipped with their (different) tangent vectors. In a differentiable point of $\partial \Omega$ one can start in directions $\beta \in [-\pi/2, \pi/2]$. In singular points this interval can be different; in our case at $r = (0,0)$ we can go in any direction. Points starting with $v_y > 0$ are attached to $\phi = \pi$, the ones with $v_y < 0$ belong to $\phi = -\pi$. The unique point with $r = (0,0)$, $v = (1,0)$ in phase space can be assigned either of the two coordinates $(4,1)$ or $(-4,-1)$ in $\mathcal{P}$. This is only a coordinate singularity and $P$ correctly maps both points onto the same point $\xi = (0,0)$, where the images of $\mathcal{S}_\pm$ meet. Strictly speaking the map $P$ is defined on a rectangle with two opposite corners identified.

Two other special lines in $\mathcal{P}$ are its upper and lower boundary

$$\mathcal{F}_\pm = \{\xi \in \mathcal{P} \,|\, p = \pm 1, s \neq \pm 4\}. \tag{12}$$

Note that $\mathcal{F}_\pm$ are half open intervals. $\mathcal{F}_\pm$ defines starting points outside of the singularity with a velocity parallel to $T$, i.e. $\beta = \pm \pi/2$, $p = \pm 1$. All the points of $\mathcal{F}_\pm$ are fixed points of the map, although physically they correspond to the above mentioned sliding motion. In a physical sense all of them are images of the two points $(\pm 4, \mp 1)$ where $\mathcal{F}$, $\mathcal{S}$ and $\Gamma^{-1}$ intersect (omitting the index $\pm$ refers to both lines).

Reversing the velocity on $\Gamma^{-1}$ we define the line

$$\Gamma = \{\xi \in \mathcal{P} \,|\, p = s/4\}, \tag{13}$$

which is the set of initial conditions that will right away hit the singularity. Therefore $\Gamma^{-1}$ is the mirror image of $\Gamma$ with respect to the $s$-axis. $\Gamma$ is of utmost importance, because this line turns out to be 1) the discontinuity of the map, 2) the boundary between our primary symbol regions and 3) the origin of the pruning front. $\Gamma$ separates two regions $A$ and $B$ in $\mathcal{P}$ (see fig. 2), where $A$ is below $\Gamma$ and $B$ above. We consider $A$ and $B$ as open sets, i.e. without the lines $\mathcal{F}_\pm$, $\mathcal{S}_\pm$ and $\Gamma$.

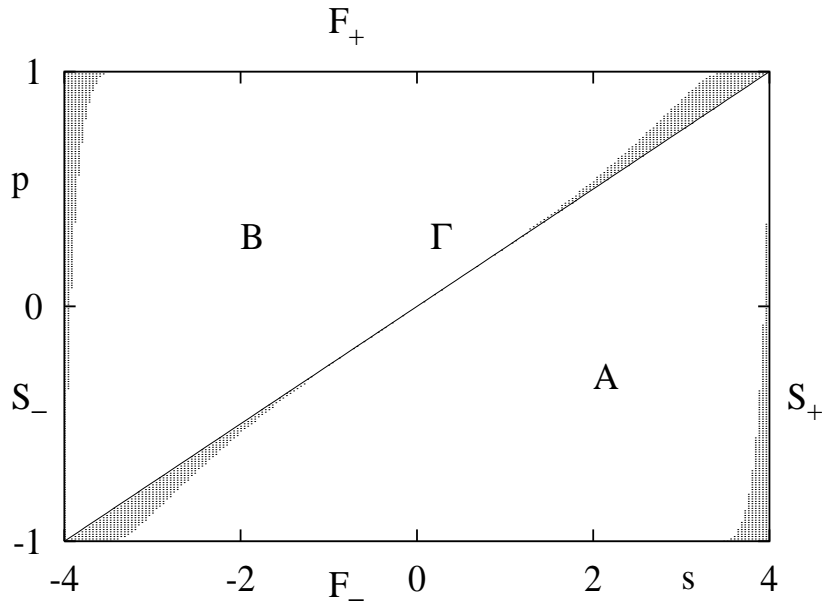

Figure 2: The regions $A$ and $B$ separated by $\Gamma$ in the Poincaré surface of section together with the fix lines $\mathcal{F}_\pm$ and the singularity lines $\mathcal{S}_\pm$. The shaded regions correspond to four intersections of the line $r + tv$ with the boundary $\partial\Omega$, i.e. three real solutions of the polynomial $Q(t)$. In the white regions there always is only one real solution.

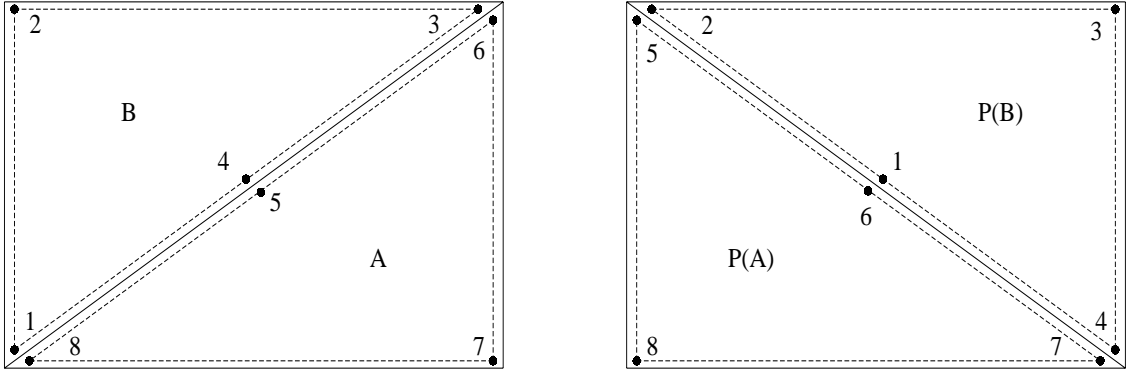

Figure 3: Demonstration of the kneading property of the billiard map. The left picture shows the regions $A$ and $B$ which are mapped by $P$ to give the picture on the right hand side. $P$ "kneads" the triangle $B$ by sliding the point 4 to the lower right corner, while in turn the point 1 is taken to the middle. The remaining points 2 and 3 stay fixed. Note that the former center 4 becomes a corner and vice versa for 1. In a similar way $A$ is deformed under $P$, by taking 5 to the upper left corner, and moving 6 to the middle. The mapping $P$ thus contracts in the direction of $\Gamma$, and expands along $\Gamma^{-1}$.

In order to understand the "kneading properties" of $P$ we need to know the behaviour of $P$ near $\Gamma$. Since $P$ is discontinuous on $\Gamma$ there are two different continuations of $P$ on $\Gamma$. Approaching $\Gamma$ from a region with negative discriminant of $Q(t)$ (shaded regions in fig. 2) the limit orbits hit the singularity and reflect at the horizontal limiting tangent. We denote the corresponding limit map by $P_s$ and find ($\pm$ denotes the sign of $s$)

$$P_s(s, s/4) = (\pm 4, \pm 1 - s/2) \ . \tag{14}$$

In the other case we ignore the encounter with the cusp and find

$$P_f(s, s/4) = (-(\pm 4 - s), (\pm 4 - s)/4). \tag{15}$$

In fig. 3 the kneading property is illustrated by mapping the regions $A$ and $B$. Here the lines $\overline{14}$ and $\overline{56}$ map according to $P_f$, while $\overline{85}$ and $\overline{43}$ map by $P_s$. The main tendency is an expansion along the direction of $\Gamma^{-1}$ and a contraction along the direction of $\Gamma$.

Since the map is discontinuous at $\Gamma$, the image of $\Gamma$ under the Poincaré map is not well defined. Moreover, there is no unique tangent vector in $\mathcal{S}$, because the boundary curve is not differentiable in this point. Note that we can assign the coordinate $s' = \pm 4$, but we cannot specify $p'$ for $\xi \in \Gamma$. One might be tempted to define the image of $\Gamma$ according to its limit under $P_f$, because also the corresponding tangent vector in $\mathcal{S}_\pm$ can be defined by an appropriate one sided differentiation. This is, however, misleading as we will see in the discussion of finite orbits that start and end in $\mathcal{S}$.

If we consider the desymmetrized billiard, see fig. 1, the corresponding billiard map $\tilde{P}$ defined on $\tilde{\mathcal{P}} = \{\xi = (s, p) \,|\, s \in [0, 4],\ p \in [-1, 1]\}$, is obtained by first using $P$ for a given $\tilde{\xi} \in \tilde{\mathcal{P}}$; if $\xi' \in \tilde{\mathcal{P}}$ we have $\tilde{\xi}' = \xi'$, otherwise $\tilde{\xi}' = (-s', -p')$.

In the numerical calculation of $P$ there is one difficulty (besides hitting $\mathcal{S}$ exactly) if we are close to $\Gamma$, especially around $\xi = (0, 0)$. Firstly the discriminant of $Q(t)$ changes quite rapidly and is very close to 0 because we are close to a double root of $Q(t)$. Secondly at the new point of $\partial\Omega$ we have very small $r$ and the calculation of $s'$ is susceptible to round-off errors. The reason is the unbounded curvature close to the singularity of $\partial\Omega$.

## 2.2 Linearized map

For an arbitrary billiard the linearized Poincaré map from $\xi_1$ to $\xi_2$ can be expressed as (see, e.g. [13])

$$DP_{21} = \begin{pmatrix} 1/n_2 & 0 \\ 0 & n_2 \end{pmatrix} \begin{pmatrix} 1 & 0 \\ -\kappa_2/n_2 & 1 \end{pmatrix} \begin{pmatrix} -1 & -l \\ 0 & -1 \end{pmatrix} \begin{pmatrix} 1 & 0 \\ -\kappa_1/n_1 & 1 \end{pmatrix} \begin{pmatrix} n_1 & 0 \\ 0 & 1/n_1 \end{pmatrix}. \quad (16)$$

Here $\kappa_i$ and $n_i$ denote the curvature and normal component, respectively, and $l$ is the geometric length between the two reflections. Lower indices indicate that the corresponding quantity is to be evaluated at $s_i$. Double indices for quantities depending on both points are usually omitted. For a periodic orbit this reduces to the more familiar form for the monodromy matrix $M$ (see, e.g. [14])

$$M = \prod_i \begin{pmatrix} 1 & l_{i,i+1} \\ 0 & 1 \end{pmatrix} \begin{pmatrix} -1 & 0 \\ 2\kappa_i/n_i & -1 \end{pmatrix}. \quad (17)$$

We prefer to use (16) because it has a special form which allows us to show that every periodic orbit (except on $\mathcal{F}$) has positive trace. The argument is along the lines of Wojtkowski's pioneering work [3], but we will use a slightly different form due to Wittek [15], who applied it to the wedge billiard. The key point is that $DP_{21}$ always has the following checker board structure (excluding $\mathcal{F}$)

$$\begin{pmatrix} + & - \\ - & + \end{pmatrix}. \quad (18)$$

Note that the product of two checker board matrices is again a checker board matrix. Let us rewrite (16) using $k_i = l\kappa_i/n_i$ as

$$DP_{21} = \begin{pmatrix} 1/n_2 & 0 \\ 0 & n_2 \end{pmatrix} \begin{pmatrix} k_1 - 1 & -l \\ (k_1 + k_2 - k_1 k_2)/l & k_2 - 1 \end{pmatrix} \begin{pmatrix} n_1 & 0 \\ 0 & 1/n_1 \end{pmatrix}. \quad (19)$$

Since $n_i \geq 0$ and $l > 0$ (excluding $\mathcal{F}_\pm$) we need to show $k_i > 1$ and

$$\frac{1}{k_1} + \frac{1}{k_2} < 1. \quad (20)$$

Because of $\kappa_i > 0$ we have $k_i > 0$ and therefore the latter inequality implies $k_i > 1$. Evaluating the trace with these inequalities we find that $\text{Tr}\, DP_{21} \geq 2$, where equality holds for $k_1 = k_2 = 1$. Starting from the explicit description of the Poincaré map in the previous section it seems quite hard to obtain this statement because the solution of a cubic equation is involved. Therefore we now look at the generating function of our map (see [16] for a review), which is just the length between successive reflections at $r(\phi_1)$ and $r(\phi_2)$. Note that in order to obtain the area preserving map $P$ in coordinates $\xi = (s, p)$ we should parameterize by $s_1$ and $s_2$. However, the calculations are more conveniently done with $l(\phi_1, \phi_2)$. Denote by $L$ the vector joining the two reflection points,

$$L(\phi_1, \phi_2) = r(\phi_2) - r(\phi_1) \quad (21)$$
$$l(\phi_1, \phi_2) = |L(\phi_1, \phi_2)| \quad (22)$$
$$= 2\sqrt{\sin^2 y (\cos^2 y + 2\cos x \cos y + 1)} \quad (23)$$

where $x = (\phi_1 + \phi_2)/2$ and $y = (\phi_1 - \phi_2)/2$. The unit velocity is given by $v = L/l$ and we obtain

$$n_1 = \langle N_1, v \rangle \qquad n_2 = -\langle N_2, v \rangle, \tag{24}$$

$$k_i = \pm l^2 \kappa_i / \langle N_i, L \rangle \tag{25}$$

$$= 3 \frac{\cos^2 y + 2\cos x \cos y + 1}{2\cos^2 y + 3\cos x \cos y + 1 \pm \sin x \sin y}, \tag{26}$$

where $+$ and $-$ corresponds to $k_1$ and $k_2$, respectively. Finally

$$\frac{1}{k_1} + \frac{1}{k_2} = \frac{2}{3} \frac{2\cos^2 y + 3\cos x \cos y + 1}{\cos^2 y + 2\cos x \cos y + 1} \tag{27}$$

$$= 1 - \frac{1}{3} \frac{\sin^2 y}{\cos^2 y + 2\cos x \cos y + 1} \tag{28}$$

$$= 1 - \frac{4}{3} \sin^4 y / l^2 < 1 \tag{29}$$

proves the inequality (20). The geometric origin of this relation is the convex scattering property [3] of the cardioid, i.e. $\partial^2 \kappa / \partial s^2 > 0$. There are a some very important consequences:

- The eigenvalues of the monodromy matrix of periodic orbits are always positive, i.e. all periodic orbits are direct hyperbolic. We conclude that periodic orbits are *maxima* of the Lagrangian $\mathcal{L} = \sum l(\phi_{i-1}, \phi_i)$ [17, 16]. Thus the numerical search for periodic orbits is very much simplified, because we do not have to find saddle points of $\mathcal{L}$, which would correspond to inverse hyperbolic orbits (see [18] for related results).

- The maximum number of conjugate points along a periodic orbit is given by the number of reflections. The reason for this is contained in the optical interpretation of (20) already described by Wojtkowski [3]: there is enough time between successive reflections in order for a conjugate point to occur. Since for a free motion there cannot be more than one conjugate point, the above statement follows.

- We can obtain an analytical estimate (from below) of the maximum Lyapunov exponent and therefore also for the Kolmogorov-Sinai entropy by a theorem from [3], see sec. 5.

Note that the first statement only holds in the *non* symmetry reduced system. This is one of the reasons why we think that for the cardioid it is worthwhile to study the non-reduced system. Furthermore notice that the convex scattering property, which is the basis for all of the above, does not hold for any other member of the family of billiards (1), which are therefore much more difficult.

## 2.3 Symmetries

The time reversal symmetry of a billiard combined with the spatial symmetry of the cardioid gives us a number of symmetry classes of orbits. We will not pass to a desymmetrized billiard map but instead do the symmetry reduction on the level of the symbolic dynamics. Here we discuss the manifestation of the symmetries in the map. In the following sections this will be translated into symbol plane.

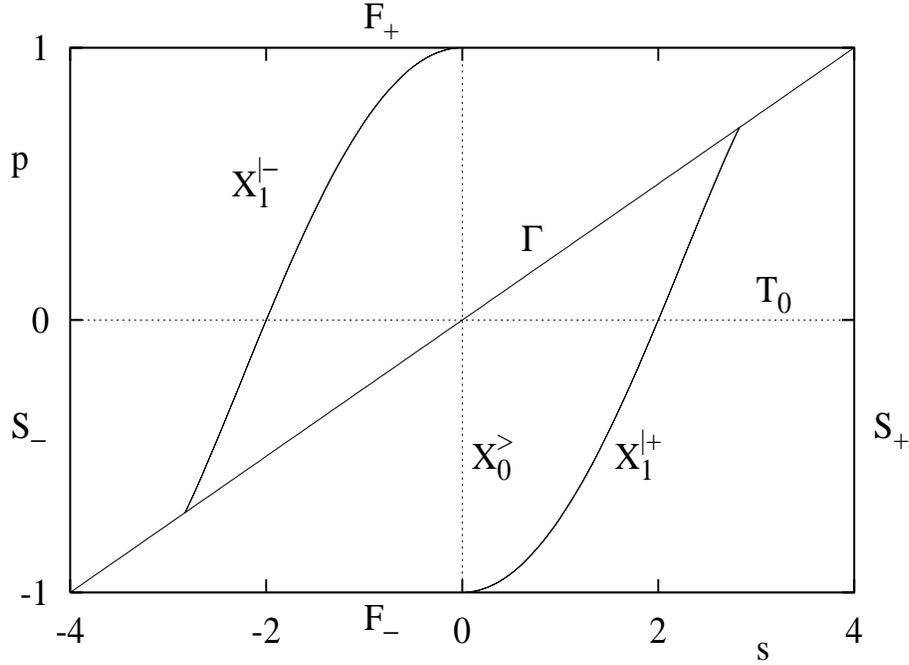

Figure 4: The symmetry lines in the Poincaré section $\mathcal{P}$. $\mathcal{T}_0$ corresponds to orbits starting at right angles on the boundary, $\mathcal{X}_0^>$ contains all orbits starting in $s = 0$, and $\mathcal{X}_1^{|\pm}$ corresponds to orbits intersecting the $x$-axis at right angles.

The standard time reversal symmetry is given by

$$T : (s, p) \mapsto (s, -p) \ . \tag{30}$$

As usual the $T$-symmetry allows for a simple expression of the inverse map as $P^{-1} = TPT$. The cardioid billiard also has reflection symmetry with respect to the $x$-axis

$$X : (s, p) \mapsto (-s, p). \tag{31}$$

$X$ and $T$ are involutions, i.e. $T^2 = id$, $X^2 = id$ and $\det T = -1$, $\det X = -1$.

Furthermore we have $P^{-n} = XP^nX$ and $P^{-n} = TP^nT$, and thus we define two families of involutions

$$\begin{aligned}
T_n &= P^nT, & T_n^2 &= \text{id}, & \det T_n &= -1, & n &= \pm 1, \pm 2 \ldots & (32)\\
X_n &= P^nX, & X_n^2 &= \text{id}, & \det X_n &= -1, & n &= \pm 1, \pm 2 \ldots & (33)
\end{aligned}$$

The fixed point sets of involutive symmetries, the so-called symmetry lines, are useful in finding symmetric periodic orbits [19] and in their classification [20]. We define

$$\begin{aligned}
\mathcal{T}_n &= \{\xi \,|\, T_n\xi = \xi\} & (34)\\
\mathcal{X}_n &= \{\xi \,|\, X_n\xi = \xi\} \ . & (35)
\end{aligned}$$

The set $\mathcal{T}_0$ contains all the orbits starting at right angles on the boundary, i.e.

$$\mathcal{T}_0 = \{\xi \in \mathcal{P} \,|\, p = 0\}. \tag{36}$$

$\mathcal{X}_0$ contains all orbits starting at the rightmost point of the cardioid with arbitrary angle

$$\mathcal{X}_0 = \{\xi \in \mathcal{P} \,|\, s = 0\}. \tag{37}$$

All iterates of the symmetry lines can be obtained by iterating $\mathcal{T}_0$, $\mathcal{T}_1$, $\mathcal{X}_0$ and $\mathcal{X}_1$, because [19]

$$\begin{array}{rclrcl} \mathcal{T}_{2n} & = & P^n \mathcal{T}_0 & \mathcal{T}_{2n+1} & = & P^n \mathcal{T}_1 \\ \mathcal{X}_{2n} & = & P^n \mathcal{X}_0 & \mathcal{X}_{2n+1} & = & P^n \mathcal{X}_1 \end{array}. \tag{38}$$

The equations $PT\xi = \xi$ for $\mathcal{T}_1$ do not have a solution in billiards without potential and therefore also $\mathcal{T}_{2n+1} = \emptyset$.

The symmetry line $\mathcal{X}_1 = \{\xi \,|\, PX\xi = \xi : (-s', p') = (s, p)\}$ corresponds to orbits that intersect the $x$-axis at right angles. Therefore we introduce the suggestive notation

$$\mathcal{X}^{>}_{2n} \equiv \mathcal{X}_{2n} \tag{39}$$

$$\mathcal{X}^{|}_{2n+1} \equiv \mathcal{X}_{2n+1} \tag{40}$$

related to the geometric form of the corresponding periodic orbits. $\mathcal{X}^{|}_1$ is obtained by using (8) which yields the condition $\sin(3\phi + \beta) = \sin(\beta)$ and finally

$$\mathcal{X}^{|+}_1 = \{\xi \,|\, p = -\cos(3\phi/2), \phi \in [0, \pi/2]\} \tag{41}$$

$$\mathcal{X}^{|-}_1 = \{\xi \,|\, p = \cos(3\phi/2), \phi \in [-\pi/2, 0]\}, \tag{42}$$

because for $|\phi| > \pi/2$ we cannot cross the $x$-axis.

If we consider the family of billiards (1) the set $\mathcal{S}$ does become a symmetry line. In the spirit of this notation we should call it $\mathcal{X}^{<}_0$. Moreover, the image respectively preimage of $\Gamma = \mathcal{X}^{<}_{-2}$ respectively $\Gamma^{-1} = \mathcal{X}^{<}_2$ is well defined in this case. The interpretation of the line of discontinuity $\Gamma$ as a symmetry line leads to a nice interpretation of finite orbits starting and ending in $\mathcal{S}$.

At intersections $\mathcal{X}_n \cap \mathcal{X}_m$, $\mathcal{X}_n \cap \mathcal{T}_m$ and $\mathcal{T}_n \cap \mathcal{T}_m$ of symmetry lines there are periodic orbits. Intersections of symmetry lines of the same type $\mathcal{X}$ or $\mathcal{T}$ are periodic orbits with (not necessarily primitive) period $|m-n|$ [19]. If $\mathcal{X}_m$ intersects $\mathcal{T}_n$ we instead have period $|2n-2m|$. This is easily seen in the following way: We have $P^n T\xi = \xi$ and $P^m X\xi = \xi$, thus one has $P^{n-m}TX\xi = \xi$ which in turn implies $P^{2(n-m)}\xi = \xi$. After symmetry reduction we always find period $|m - n|$.

Let us now study the the intersections of symmetry lines visible in fig. 4. Already the basic lines $\mathcal{T}_0$ and $\mathcal{X}^{>}_0$ intersect at $\xi = (0,0)$, but the corresponding orbit is not periodic but instead a finite orbit running along the $x$-axis. The point $\xi = (0,0)$ is also an intersection of $\Gamma$ and $\Gamma^{-1}$, i.e. the corresponding orbit starts on $\mathcal{S}$, is mapped to $\Gamma^{-1}$ *and* $\Gamma$ and back into $\mathcal{S}$ (with undefined $p'$).

The periodic orbit with the shortest period (besides all the parabolic fixed points on $\mathcal{F}_\pm$) is given by the intersection of $\mathcal{T}_0$ and $\mathcal{X}^{|\pm}_1$ at $(\pm 2, 0)$ respectively $\phi = \pm \pi/3$. There is one more intersection in fig. 4: $\mathcal{X}^{|}_1$ has endpoints on $\Gamma$. The (well defined) preimage of this point on $\Gamma$ is on $\Gamma^{-1}$, so that this orbit has two reflections besides the point in the singularity. A discussion of symmetric orbits with higher period will be postponed until we have the symbolic dynamics at hand.

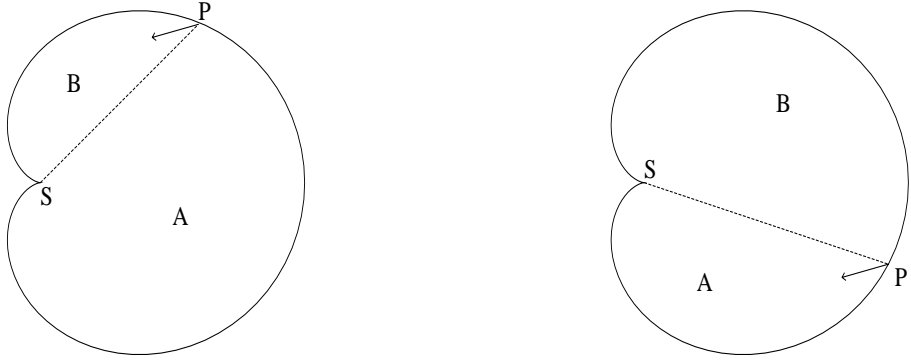

Figure 5: Examples for the determination of the symbols $A$ and $B$ in position space by the following rule: Connect the singularity $S$ with the current point $P$. Now determine if the velocity vector is inside the sector formed by $\vec{PS}$ and the oriented tangent vector. If the velocity vector is inside the sector then $B$ is assigned, otherwise $A$. In the left example, the symbol $B$ is associated, whereas in the right example, the corresponding symbol is $A$.

## 3 Symbolic Dynamics

The discussion of the Poincaré map and its discontinuities proposes a natural choice for the initial partition of $\mathcal{P}$: the regions $A$ and $B$ as separated by $\Gamma$. For a given velocity on $\partial\Omega$ one can easily read off its symbol as illustrated in fig. 5. Transforming this description in the Poincaré section to the one in configuration space means to consider two consecutive points $\phi$ and $\phi'$ on the boundary. If $\phi' > \phi$ the letter $B$ is assigned for $\phi$, if $\phi' < \phi$ we obtain $A$. Furthermore we exclude the cases $\phi = \phi'$, $\phi = \pm\pi$ and $\phi' = \pm\pi$ in order to ensure that the regions $A$ and $B$ correspond to open sets in $\mathcal{P}$.

The construction of the partition is as follows: If we superimpose $\Gamma$ and $\Gamma^{-1}$ we obtain four cells in $\mathcal{P}$ labeled $A.A$, $A.B$, $B.B$, and $B.A$ which are shown in fig. 6. The forward image of these cells generates "past" stripes $AA.$, $BA.$, $AB.$ and $BB.$ basically along the direction of $\Gamma^{-1}$ ordered by increasing $p$. The operation of $P$ on a symbol sequence (word) just shifts the dot to the right. The sequence $AB.$ for a strip means that its preimage is in $.B$ and moreover in that part of $.B$ whose preimage is in $.A$.

The backward images of the initial cells $A.A$, $A.B$, $B.A$, and $B.B$ give stripes elongated in the direction of $\Gamma$ labeled $.AA$, $.AB$, $.BA$ and $.BB$, which are just the images of the past stripes under $T$. The new words $.AA$, $.AB$, $.BA$ and $.BB$ tell about the future of a given strip. The intersection of these two sets of stripes generates a partition of $\mathcal{P}$ into $2^4$ cells which is also displayed in fig. 6; each cell is uniquely labeled by four symbols. Iterating this process gives a finer and finer subdivision, where the hope is that the cell size eventually goes to zero so that every point in $\mathcal{P}$ can be uniquely labeled by a doubly infinite sequence. We cannot prove that this process generates a partition, but we have good numerical evidence that it does.

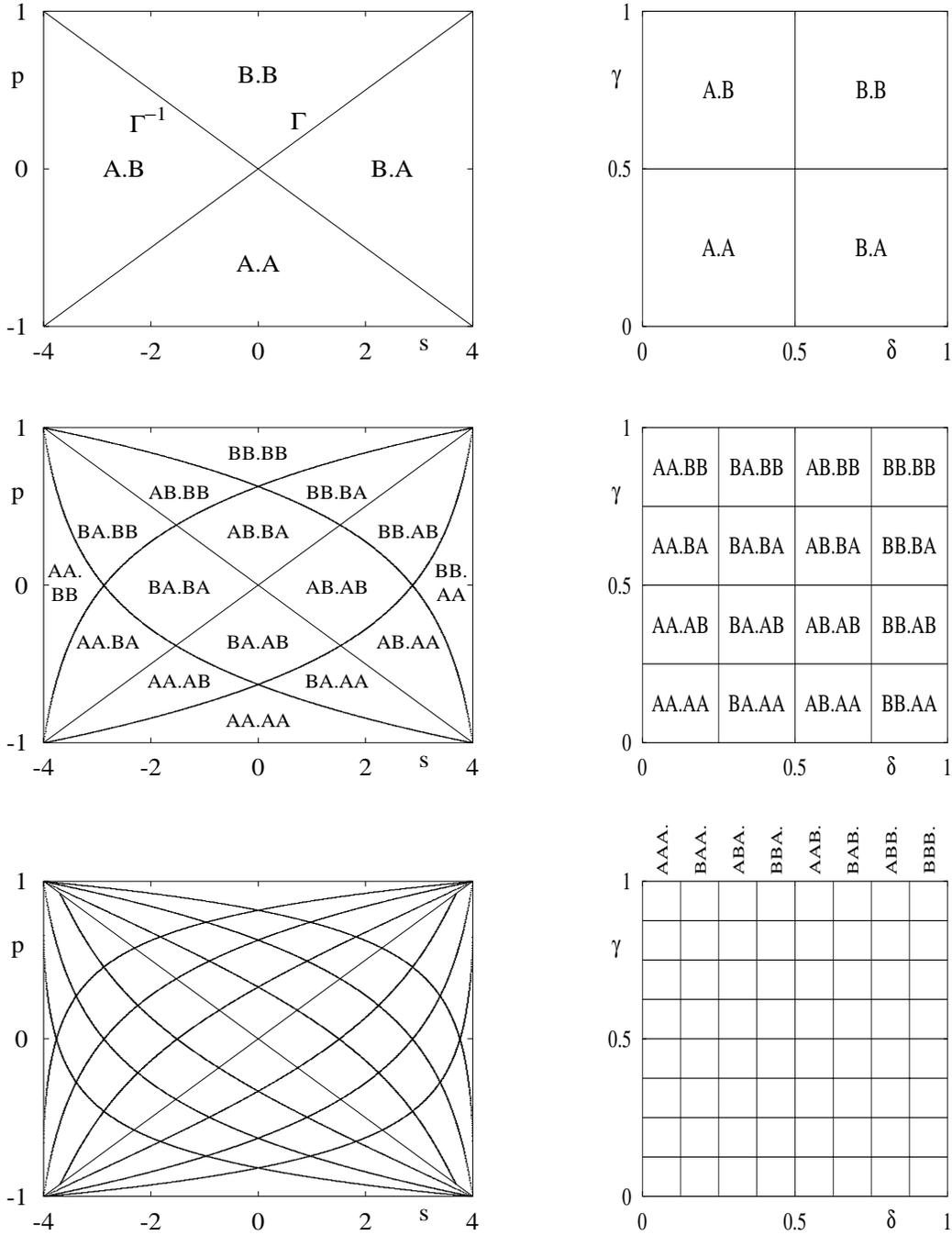

Figure 6: The division of $\mathcal{P}$ by preimages of $\Gamma$ and images of $\Gamma^{-1}$ into $2^2$ cells with 2 symbols, $2^4$ cells with 4 symbols and $2^6$ cells with 6 symbols. On the right hand side the corresponding division of the symbol plane is shown. Every line has 1, 3 respectively 7 intersections with other lines.

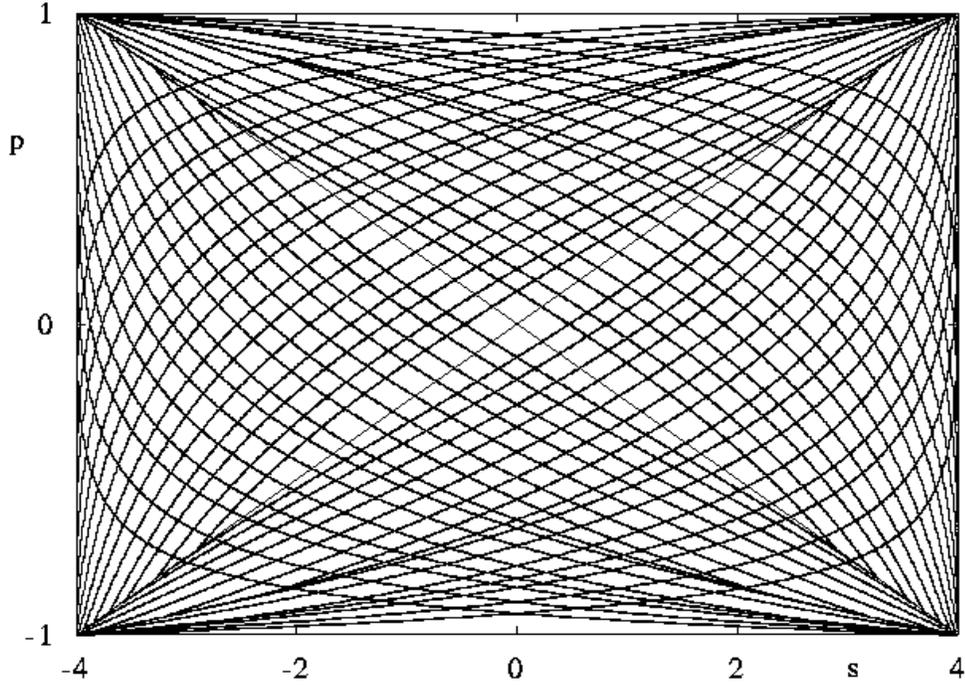

Figure 7: The division of $\mathcal{P}$ into $2^{10} - 4$ cells with 10 symbols. At this iteration level four lines have less than 31 intersections, indicating the onset of pruning.

## 3.1 Symbol plane

For any word $\alpha.\omega = \ldots s_{-2}s_{-1}.s_1s_2\ldots$ a point $(\delta, \gamma) \in [0,1] \times [0,1]$ in the symbol plane for the "future"-coordinate $\gamma$ is calculated by

$$\gamma = \sum_{i=1}^{\infty} s_i 2^{-i} \; , \tag{43}$$

and for the "past"-coordinate by

$$\delta = \sum_{i=1}^{\infty} s_{-i} 2^{-i} \; , \tag{44}$$

where in the context of numerical interpretation $s_i$ is zero (or one) for symbol $A$ (or $B$).

A quite surprising observation is that the ordering of stripes in the Poincaré section corresponds to the ordering of words $(\delta, \gamma)$ in the symbol plane. Thus the symbolic dynamics is already well-ordered [2, 21]. The dynamics in the symbol plane is a shift on the symbol sequences, i.e. it maps according to the baker map (see, e.g. [22]).

Fig. 7 shows the onset of pruning. For symbol length 8 we have $2^8$ cells, i.e. the 16 past and 16 future stripes intersect pairwise. At symbol length 10 this is not true any more. The future stripe $.AB^4$, which is the first one below $\Gamma$ does not intersect the two outermost past stripes $B^5.$, so that there is no cell e.g. with label $B^5.AB^4$ (see fig. 8 for a magnification of the relevant region in $\mathcal{P}$). In the magnification it is also visible, that the future stripe $.BA^4$ above $\Gamma$ still intersects the past stripe $B^5.$, such that the cell $B^5.BA^4$ exists.

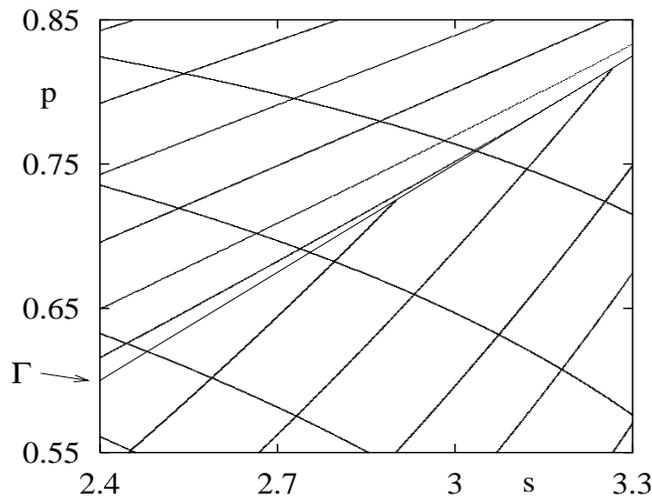

Figure 8: Magnification of fig. 7. The upper right stripe is $B^5.$, the stripe above $\Gamma$ is $.BA^4$, and $.AB^4$ is the one below. Since the stripes $B^5.$ and $.AB^4$ do not intersect, the sequence $B^5.AB^4$ is forbidden, whereas $.BA^4$ and $B^5.$ still intersect, such that $B^5.BA^4$ is allowed.

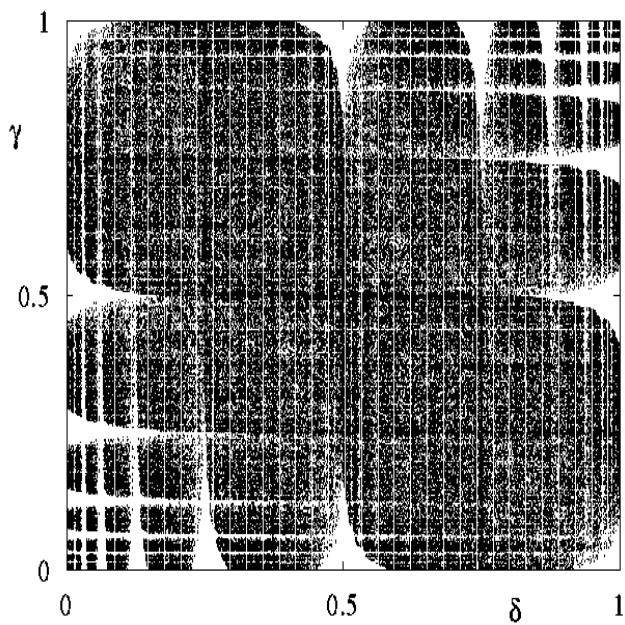

Figure 9: $10^6$ iterations of a single orbit shown in the symbol plane truncated to word length $2 \times 20$.

To find out systematically which cells are forbidden we plot the truncated symbolic past and future for points from a generic trajectory in the symbol plane (see, e.g. [21]). The result is shown in fig. 9. One observes that the picture is symmetric under reflections with respect to the diagonals. This is caused by the underlying symmetries $T$, $X$ and $TX$: For a word $\alpha.\omega$ the symmetry operations are realized by the operations of reading backwards $\overleftarrow{\omega}.\overleftarrow{\alpha}$ and taking the complement $\hat{\alpha}.\hat{\omega}$. $T$ is realized by $\overset{\frown}{\omega}.\overset{\frown}{\alpha}$, and $X$ is represented by $\overset{\leftarrow}{\omega}.\overset{\leftarrow}{\alpha}$ alone. Thus $TX$ is represented by $\hat{\alpha}.\hat{\omega}$. Using eqs. (43) and (44) one can easily calculate the action of these operations in the symbol plane. For a given word $\alpha.\omega$ with coordinates $(\delta, \gamma)$ in the symbol plane we have

$$
\begin{aligned}
T\alpha.\omega &= \hat{\overset{\frown}{\omega}}.\hat{\overset{\frown}{\alpha}} & T(\delta, \gamma) &= (1 - \gamma, 1 - \delta) \\
X\alpha.\omega &= \overset{\leftarrow}{\omega}.\overset{\leftarrow}{\alpha} & X(\delta, \gamma) &= (\gamma, \delta) \\
TX\alpha.\omega &= \hat{\alpha}.\hat{\omega} & TX(\delta, \gamma) &= (1 - \delta, 1 - \gamma) \;.
\end{aligned}
\qquad (45)
$$

Similarly we now can define the basic symmetry lines for the symbol plane. $\mathcal{X}_0$ is given by the diagonal $\delta = \gamma$ and $\mathcal{T}_0$ is the other diagonal $\delta + \gamma = 1$. A detailed discussion of the symmetries in connection with periodic orbits will be done in sec. 4.

## 3.2 Pruning front

Returning to the onset of pruning at word length 10 close to $\Gamma$ one expects from the geometry of the billiard that the origin of pruning is the singularity of the cardioid. Therefore, we now study the set of symbol sequences of orbits that almost hit the singularity. Two possibilities arise: on the one hand the orbit may just miss the singularity, or on the other hand, it may be reflected very close to the singularity. The limiting cases of these types of orbits generate the pruning fronts [2, 21] between allowed and forbidden sequences, see fig. 10.

We think that this pruning front fulfills the conjecture stated in [2] that the region enclosed by the front and the $\delta$-axis specifies the primary pruned region in the symbol plane. All the other forbidden cells visible in fig. 9 are related via the symmetry operations or they are images or preimages of these regions. We cannot prove that the pruning front separates allowed and forbidden orbits in the symbol plane, but we have found no counterexample.

The two pruning fronts are obtained from the two possible limiting maps $P_f$ and $P_s$ on $\Gamma$. The recipe is to start with $\xi \in \Gamma$ and to use either $P_f$ or $P_s$ to map across the discontinuity. The corresponding symbols can be read off from figures 2 and 3; if we start for example close to the line $\overline{56}$ shown in fig. 3, the current symbol is $A$ and the application of $P_f$ carries us into $B$. In the case of $\xi' \in \mathcal{S}$ we assign $A$ if $s' = 4$ and $B$ if $s' = -4$ for the reflection in the singularity. After this crucial step all further forward images under $P$ are well defined, similarly the preimages of $\xi \in \Gamma$. Therefore an infinite symbolic past and future can be assigned to $\xi$ and the corresponding point in the symbol plane is part of the $f$- or $s$-pruning front, depending on the initial mapping step. From the two pruning fronts in fig. 10 the left with $\delta < 0.5$ is generated using $P_s$ and will be called $s$–pruning front, and the right is obtained using $P_f$, and will be called $f$–pruning front. In the division of $\mathcal{P}$, see fig. 8, the non-existing cells below $\Gamma$ correspond to squares below the $f$–pruning front, and forbidden cells above $\Gamma$ (which occur first for a division of $\mathcal{P}$ into $2^{12}$ cells) correspond to squares below the $s$–pruning front.

As a result of the above construction the two pruning fronts are related. Denote by $\delta_s$ and $\delta_f$ the respective coordinates of the two pruning fronts for the same $\gamma$. We then find

$$\delta_f = \frac{1}{2} + 2\left(\frac{1}{2} - \delta_s\right). \qquad (46)$$

Moreover the $f$–front (and the pruned region it encloses) is invariant under $X_1$, i.e. invariant under the map $(\delta, \gamma) \mapsto (\gamma/2 + 1/2, 2\delta - 1)$. These symmetries in the front induce relations between the pruned words listed in table 1.

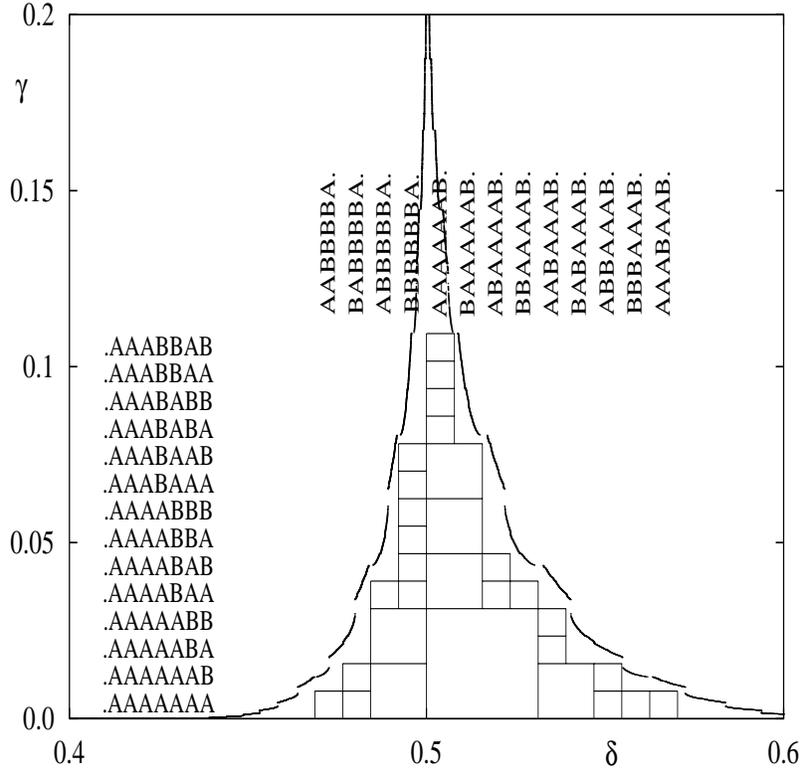

Figure 10: Magnification of the symbol plane around the primary pruning region. The $s$–pruning front left of $\delta = 0.5$ and the $f$–pruning front to the right separate allowed and forbidden code sequences. The squares of length $2^{-5}$, $2^{-6}$ and $2^{-7}$ correspond to forbidden words of length 10, 12 and 14, respectively. From this one can construct the words in table 1. The shown part of the symbol plane corresponds to the lower right corner of the Poincaré section. The primary pruning region extends to $(0, 0.375)$ to the left, to $(0, 0.75)$ to the right and up to $(0.5, 0.5)$ to the top.

A complete list of the primary forbidden sequences can therefore be read off from fig. 10, where also the corresponding squares with length $2^{-5}$, $2^{-6}$ and $2^{-7}$ are shown. The pruned words of width 10,...,14 are given in table 1.

Looking at fig. 9, one observes that the density of points is much higher close to the $f$–pruning front, while it is low close to the $s$–pruning front (this is similar to the three-disk billiard [21]). The reason is that the mapping from the Poincaré section to the symbol plane is not area preserving, e.g. the small cells just above $\Gamma$ shown in fig. 8 and the much larger ones below $\Gamma$ are represented in the symbol plane by squares of the same size. Considering a generic trajectory which fills the Poincaré section uniformly there is only a low probability in the symbol plane for squares whose corresponding cell in the Poincaré section is small.

# 4 Periodic Orbits

Periodic orbits with period $n$ are fixed points of $P^n$, i.e. $P^n \xi = \xi$. Thus the corresponding symbol sequence is also periodic and will be denoted by $\overline{\omega}$, where $\omega$ is a word consisting of $n$ letters $\omega_i$. The same periodic orbit can be denoted by $n$ different words all obtained from one word by a cyclic permutation of the letters, e.g. $\overline{ABB} \equiv \overline{BBA} \equiv \overline{BAB}$. Therefore with every periodic orbit a class of code words is associated. Two words $\overline{\omega}$ and $\overline{\omega}'$ belong to the same cyclic class if they can be obtained from each other by a cyclic permutation. Periodic orbits, which are not multiple traversals of an underlying shorter periodic orbit are called primitive periodic orbits.

| word length | pruned words $\omega$ | | | |
|---|---|---|---|---|
| 10 | $A^4B.A^5$ | | | |
| 11 | $B^5A.A^5$ | $A^5B.A^4B$ | | |
| 12 | $A^5B.A^3BA^2$ | $ABA^3B.A^6$ | | |
| 13 | $AB^4A.A^7$ | $BAB^4A.A^6$ | $B^2A^3B.A^7$ | $AB^2A^3B.A^6$ |
| | $A^2BA^3B.A^5B$ | $B^5A.A^4BA^2$ | $B^6A.A^3BA^2$ | $BA^4B.A^4BA^2$ |
| | $ABA^4B.A^4BA$ | $A^6B.A^3BAB$ | $A^6B.A^3BBA$ | |
| 14 | $A^3.BA^2B.A^7$ | | | |

Table 1: Pruned words in the cardioid billiard up to length 14. The shown words $\alpha.\omega$ and additionally $\overleftarrow{\omega}.\overleftarrow{\alpha}$, $\hat{\alpha}.\hat{\omega}$ and $\hat{\overleftarrow{\alpha}}.\hat{\overleftarrow{\omega}}$ are forbidden. Code words which are already pruned by shorter words are omitted. Notice, that for example the squares $B^5A.A^6$ and $B^5A.A^5B$ of length 12 shown in fig. 10 rule out all words of the form $B^5A.A^5$ of length 11.

For a periodic symbol sequence $\overline{\omega} = \overline{\omega_1 \ldots \omega_n}$, the corresponding point in the symbol plane $(\delta, \gamma)$ is given by

$$\gamma = \frac{N(\omega)}{2^n - 1} = \frac{1}{1 - 2^{-n}} \sum_{i=1}^{n} \omega_i \, 2^{-i} \tag{47}$$

$$\delta = \frac{N(\overleftarrow{\omega})}{2^n - 1} = \frac{1}{1 - 2^{-n}} \sum_{i=1}^{n} \omega_{-i+n+1} \, 2^{-i} \quad, \tag{48}$$

where $N(\omega) = \sum_{i=1}^{n} \omega_i 2^{n-i}$ is the binary interpretation of $\omega$. We choose the symbol sequence with the smallest value $\gamma$ as a representative for that orbit.

The classical properties length, stability and Maslov index of periodic orbits belonging to different cyclic classes, which are related by $X$, $T$ or $XT$, are identical. Therefore different cyclic classes will be combined into symmetry classes if their code words can be transformed into each other by reflection and/or time-reversal. Thus it is sufficient to calculate only one periodic orbit for every symmetry class. As symmetry classes we define

$$\begin{aligned} C_{\text{full}} &= \{\overline{\omega} \mid \overline{\omega} \equiv \overline{\overleftarrow{\omega}} \equiv \overline{\hat{\overleftarrow{\omega}}} \equiv \overline{\hat{\omega}}\} \\ C_{\text{TX}} &= \{\overline{\omega} \mid \overline{\omega} \equiv \overline{\hat{\omega}} \neq \overline{\overleftarrow{\omega}}\} \\ C_{\text{X}} &= \{\overline{\omega} \mid \overline{\omega} \equiv \overline{\overleftarrow{\omega}} \neq \overline{\hat{\omega}}\} \\ C_{\text{T}} &= \{\overline{\omega} \mid \overline{\omega} \equiv \overline{\hat{\overleftarrow{\omega}}} \neq \overline{\overleftarrow{\omega}}\} \\ C_{\text{no}} &= \{\overline{\omega} \mid \overline{\omega} \neq \overline{\overleftarrow{\omega}}, \overline{\hat{\overleftarrow{\omega}}}, \overline{\hat{\omega}}\} \quad, \end{aligned} \tag{49}$$

using the operations defined in eq. (45) (see [14, 23] for a similar classification in the case of the hyperbola billiard). Within every symmetry class we choose the sequences with the smallest binary interpretation $N(\omega)$ as a representative for that class.

To every symmetry class a multiplicity is associated, which counts the corresponding number of different cyclic classes. For $C_{\text{no}}$ the multiplicity is 4, $C_{\text{T}}$, $C_{\text{X}}$ and $C_{\text{TX}}$ have multiplicity 2 and $C_{\text{full}}$ has multiplicity 1. For each symmetry class at least one example of a corresponding periodic orbits is shown in fig. 11.

In the case of no symmetry $\leftarrow$, $\hat{-}$ and $\hat{\ }$ give the symmetric partners, so that each of the symmetry classes is closed under these operations. If $\overline{\omega} \in C_{\text{T}}$ or $\overline{\omega} \in C_{\text{X}}$ the corresponding symmetric partners are generated by $\leftarrow$ or $\hat{-}$ respectively. For $\overline{\omega} \in C_{\text{TX}}$ one obtains the symmetric partner by $\hat{-}$ or $\leftarrow$, since $\hat{\overleftarrow{\omega}} \equiv \overleftarrow{\omega}$. Thus despite the fact that the orbit shown in fig. 11c) looks symmetric,

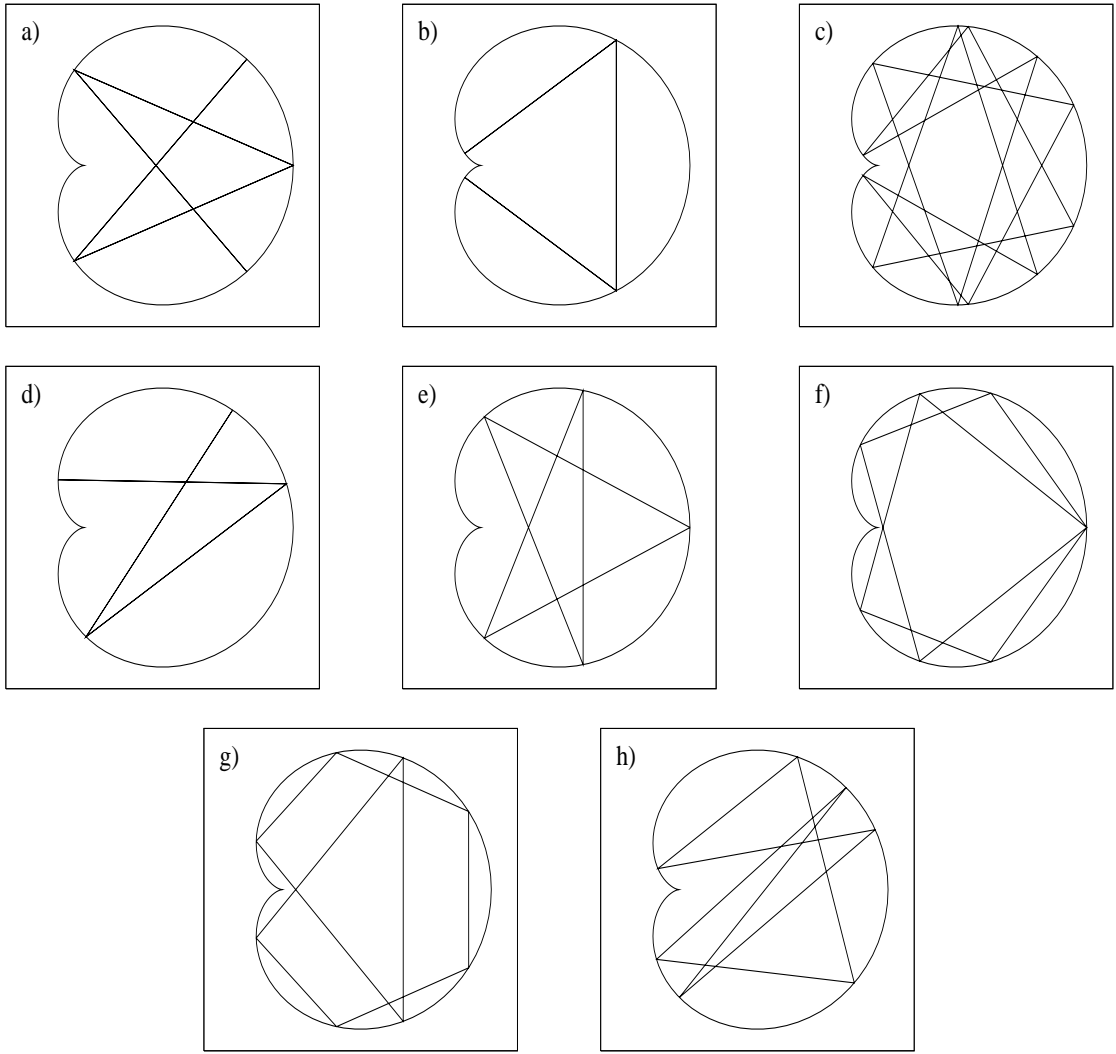

Figure 11: Examples of periodic orbits: a) $\overline{\omega} = \overline{AABABBAB} \in C_{\text{full}}^{>\perp}$, b) $\overline{\omega} = \overline{AAABBB} \in C_{\text{full}}^{|\perp}$ c) $\overline{\omega} = \overline{AAABAABBBABB} \in C_{\text{TX}}$, d) $\overline{\omega} = \overline{AABABB} \in C_{\text{T}}$, e) $\overline{\omega} = \overline{AABAB} \in C_{\text{X}}^{|>}$, f) $\overline{\omega} = \overline{AAAABAAB} \in C_{\text{X}}^{>>}$, g) $\overline{\omega} = \overline{AAAAABAB} \in C_{\text{X}}^{\|}$ and h) $\overline{\omega} = \overline{AAABABB} \in C_{\text{no}}$.

application of $X$ yields the corresponding time-reversed orbit.

A finer classification of orbits with $X$ and full symmetry is obtained by using the symmetry lines (see [20] for a similar discussion):

- $C_X^{>>} \subset C_X$ contains periodic orbits of period $2n$, which are intersections of the type $\mathcal{X}_{2n}^{>} \cap \mathcal{X}_0^{>}$. The orbit has two different points with $\phi = 0$.

- $C_X^{|>} \subset C_X$ contains periodic orbits of period $2n - 1$, which are intersections of the type $\mathcal{X}_{2n}^{>} \cap \mathcal{X}_1^{|}$. The orbit has one perpendicular intersection with the $x$-axis and one point with $\phi = 0$.

- $C_X^{||} \subset C_X$ contains periodic orbits of period $2n$, which are intersections of the type $\mathcal{X}_{2n+1}^{|} \cap \mathcal{X}_1^{|}$. The orbit has two perpendicular intersections with the $x$-axis.

- $C_{\text{full}}^{>\perp} \subset C_{\text{full}}$ contains periodic orbits of period $4n$ corresponding to intersections of the type $\mathcal{X}_{2n}^{>} \cap \mathcal{T}_0$. The orbit has two perpendicular reflections with $\partial \Omega$ and one reflection at $\phi = 0$ which is traversed in both directions.

- $C_{\text{full}}^{|\perp} \subset C_{\text{full}}$ contains periodic orbits of period $4n + 2$ corresponding to intersections of the type $\mathcal{X}_{2n+1}^{|} \cap \mathcal{T}_0$. The orbit has two perpendicular reflections with $\partial \Omega$ and one perpendicular intersection with the $x$-axis which is traversed in both directions.

Orbits with $T$ or $TX$ or full symmetry are only possible for even $n$, since $\hat{}$ exchanges the numbers of $A$ and $B$. If $n$ is odd, only orbits from $C_X^{|>}$ and $C_{\text{no}}$ exist. Elements from $C_{\text{no}}$ only exist for $n \geq 7$.

In table 2 the number of primitive cyclic classes, primitive symmetry classes and the corresponding number of elements in the different symmetry classes are shown for $n = 2, \ldots, 15$. For period one there are two periodic sequences $\overline{A}$ and $\overline{B}$, which correspond to the lines $\mathcal{F}_\pm$ in our system. The shortest hyperbolic periodic orbit has period 2 and the sequence $\overline{AB}$.

Due to the pruning not every orbit is allowed. In table 2 the number of pruned symmetry classes are indicated by $-m$ for the corresponding symmetry classes. Starting at period 5 the orbits $\overline{A^j B}$ are pruned for $j \geq 4$. All other periodic code words up to period 9 correspond to physical orbits. From period 9 the orbits $\overline{A^j B A^3 B}$, $j \geq 4$ are missing, and for period 10 in addition $\overline{A^6 B A A B}$. Three examples of forbidden orbits are shown in fig. 12.

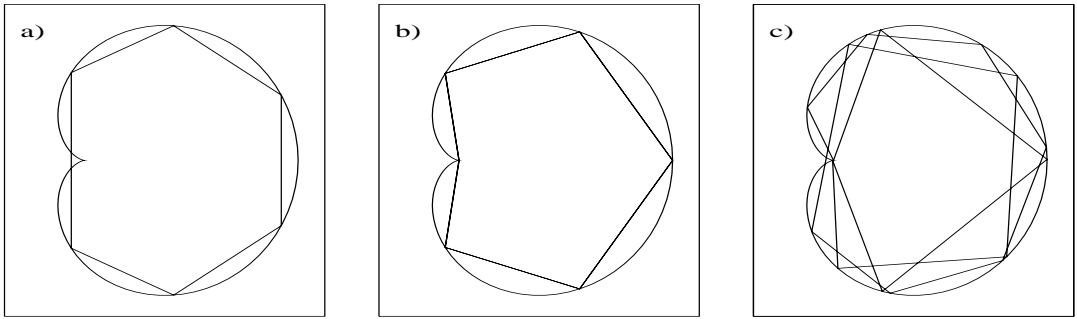

Figure 12: Examples of pruned periodic orbits: a) $f$–pruning: $\overline{A^5 B}$, b) $s$–pruning: $\overline{A^6 B^6}$ and c) combined pruning: $\overline{A^7 B A^4 B^4}$. Note that in c) the reflection in the singularity takes place only from above.

| n | cyclic classes | symmetry classes | $C_{\text{full}}$ | | $C_{\text{TX}}$ | $C_{\text{T}}$ | $C_{\text{X}}$ | | | $C_{\text{no}}$ |
|---|---|---|---|---|---|---|---|---|---|---|
| | | | $C_{\text{full}}^{|\perp}$ | $C_{\text{full}}^{>\perp}$ | | | $C_{\text{X}}^{\|}$ | $C_{\text{X}}^{>|}$ | $C_{\text{X}}^{>>}$ | |
| 2 | 1 | 1 | 1 | | | | | | | |
| 3 | 2 | 1 | | | | | | 1 | | |
| 4 | 3 | 2 | 1 | | | | 1 | | | |
| 5 | 6–2 | 3–1 | | | | | | 3–1 | | |
| 6 | 9–2 | 5–1 | 1 | | | 1 | 2–1 | | 1 | |
| 7 | 18–2 | 8–1 | | | | | | 7–1 | | 1 |
| 8 | 30–2 | 14–1 | | 2 | | 2 | 6–1 | | 2 | 2 |
| 9 | 56–4 | 21–2 | | | | | | 14–2 | | 7 |
| 10 | 99–6 | 39–3 | 3 | | | 6 | 12–2 | | 6–1 | 12 |
| 11 | 186–12 | 62–6 | | | | | | 31–6 | | 31 |
| 12 | 335–33 | 112–13 | | 3–1 | 1 | 12–1 | 27–4 | | 12–3 | 57–4 |
| 13 | 630–76 | 189–25 | | | | | | 63–12 | | 126–13 |
| 14 | 1161–145 | 352–46 | 7–1 | | 1 | 28–2 | 56–10 | | 28–4 | 232–29 |
| 15 | 2182–314 | 607–89 | | | | | | 123–21 | | 484–68 |

Table 2: Number of primitive cyclic classes, primitive symmetry classes and the number of elements in the corresponding symmetry classes for $n = 2, \ldots, 15$ in the case of a complete binary symbolic dynamics. The number $m$ of pruned orbits in the cardioid billiard is indicated by $-m$ in the corresponding classes.

Some of the forbidden orbits are obviously ruled out by the pruned symbols given in table 1. For example the pruned word $A^4 B.A^5$ rules out the existence of all the periodic orbits of the form $\overline{A^j B}$, for $j \geq 5$ but not for $j = 4$. The existence of orbits can not be deduced from the pruned words (of finite length), but only the nonexistence of periodic orbits. The reason is that a periodic orbit is an infinite sequence, which might be pruned by an extremely long word which is much longer than the period. This is nicely illustrated in the case of the periodic orbit $\overline{A^4 B}$, which is forbidden by the word $ABA^4B.A^4BA$ or $BA^4B.A^4BA^2$ of length 13, see fig. 10 and table 1.

The orbit $\overline{A^5 B}$ shown in fig. 12a) is an example of $f$–pruning. In fact, the corresponding point in the symbol plane is inside the pruned square $A^4 B.A^5$, see fig. 10. As an example for $s$–pruning in fig. 12b) the forbidden orbit $\overline{A^6 B^6}$ is shown, which is pruned by the word $B^5 A.A^5$, whose square is below the $s$–pruning front. In addition one also has orbits where combinations of both pruning mechanisms occur, see fig. 12c).

For long orbits especially the $s$–pruning is the reason for some numerical difficulties in deciding whether an orbit has a reflection next to the cusp or whether it is pruned. Some periodic orbits close to the cusp have quite large eigenvalues, e.g. for period 11 the most unstable orbit has eigenvalue 663393 and the eigenvalue for the most unstable orbit of period 15 is even larger than one million. Note that such large instabilities cannot be calculated accurately using the monodromy matrix because of round-off errors; therefore we use the method described in [17].

## 4.1 Desymmetrized cardioid billiard

The code word in the desymmetrized system can be easily read off from the code word in the full system in the following way: if two adjacent symbols are identical, assign 0, otherwise 1. For example the code for $\overline{AAAB}$ in the desymmetrized systems is $\overline{0011}$. Periodic orbits of the full system with full or $TX$ symmetry are reduced in length by a factor two. For example $\overline{AAABBB} \in C_{\text{full}}$ maps to $\overline{001001} = \overline{001}$, or $\overline{AAABAABBBABB} \in C_{TX}$ in the desymmetrized system is given by $\overline{001101001101} = \overline{001101}$.

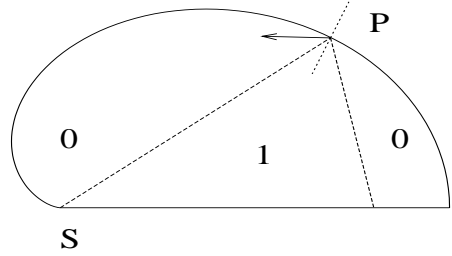

Figure 13: The rule for the symbols 0 and 1 in the desymmetrized billiard in configuration space: Connect the cusp with the current point on the boundary. Then reflect this line at the normal vector in the current point. If the actual velocity vector is in the sector formed by these two lines assign the symbol 1, otherwise 0.

Likewise one can construct the code word in the full system from the one in the desymmetrized system. Only if the code word in the desymmetrized system has an odd number of 1's, one has to double the word, because otherwise one would not obtain a periodic word. These orbits are just the ones with $C_{\text{full}}$ or $C_{TX}$ symmetry in the full system.

The multiplicities of orbits in the desymmetrized system can be determined from the symmetry of the corresponding orbit in the full system: For an orbit from $C_{\text{full}}$ the multiplicity is still 1; in the case of $C_X$ and $C_T$ the multiplicity reduces from 2 to 1. Orbits from $C_{TX}$ and $C_{\text{no}}$ are almost indistiguishable in the desymmetrized system, they both have multiplicity two. For the relation between the code length $n$ in the full and $n'$ in the desymmetrized system and the corresponding multiplicities $m$ and $m'$ the rule $\frac{n}{n'} \frac{m}{m'} = 2$ holds. If $n/n' = 2$ then also the geometric length and the stability exponent (which is the natural logarithm of the larger eigenvalue of the monodromy matrix $M$) are divided by two.

The sign of the trace of $M$ changes with every reflection at the $x$-axis. The total number of reflections at the $x$-axis is equal to the number of 1's in the code word of the desymmetrized system. To understand this fact it is helpful to interpret the symbols 0 and 1 in the reduced system.

In the reduced Poincaré section the symbol 0 corresponds to all points $\xi$ with $|p| > s/4$, and 1 to all points with $|p| < s/4$ (again excluding $\mathcal{S}$ and $\mathcal{F}$). Notice that the region corresponding to 0 is not simply connected. In configuration space this translates to the rule illustraed in fig. 13). For $s \leq \sqrt{16/3}$ all orbits from the sector 1 immediately reflect at the $x$-axis. For $s > \sqrt{16/3}$ only part of these orbits hits the $x$-axis. The other part with negative $p$ encounters a reflection with the boundary $\partial \Omega$ which gets it closer to $s = 0$. All the remaining reflections of this part get the symbol 0, i.e. they must eventually approach $s = 0$ and hit the $x$-axis before the next symbol 1 can occur.

|   | | | > | ⊥ |
|---|---|---|---|
| | | $\beta\overleftarrow{\beta}$ | $0\beta\overleftarrow{\beta}$ | $1\beta\overleftarrow{\beta}$ |
| > | $0\beta\overleftarrow{\beta}$ | $0\beta0\overleftarrow{\beta}$ | $1\beta0\overleftarrow{\beta}$ |
| ⊥ | $1\beta\overleftarrow{\beta}$ | $1\beta0\overleftarrow{\beta}$ | $1\beta1\overleftarrow{\beta}$ |

(50)

Table 3: Form of the code words for periodic orbits in the desymmetrized system.

| $C_{\text{full}}^{|\perp}$ | $C_{\text{full}}^{>\perp}$ | $C_{\text{T}}$ | $C_{\text{X}}^{\|\|}$ | $C_{\text{X}}^{>|}$ | $C_{\text{X}}^{>>}$ |
|---|---|---|---|---|---|
| $\alpha\hat{\overleftarrow{\alpha}}\hat{\alpha}\overleftarrow{\alpha}$ | $\alpha Y\hat{\overleftarrow{\alpha}}\hat{\alpha}\hat{Y}\overleftarrow{\alpha}$ | $\alpha\hat{\overleftarrow{\alpha}}$ | $Y\alpha Z\overleftarrow{\alpha}$ | $\alpha Y\overleftarrow{\alpha}$ | $\alpha\overleftarrow{\alpha}$ |

Table 4: Form of the code words for symmetric periodic orbits of the full cardioid billiard. $Y$ and $Z$ stand for an arbitrary single letter, $\alpha$ for an arbitrary word.

The fact that the number $\#_1(\omega)$ of 1's in the code word gives the number of reflections with the $x$-axis is important for calculating the Maslov index which enters into the Gutzwiller trace formula. For a billiard system the Maslov index $\nu$ is twice the number of reflections at the boundary with Dirichlet boundary condition plus the maximal number $\mu$ of conjugate points [24]. For the desymmetrized cardioid billiard we have in the case of Dirichlet boundary conditions $\nu = 2n + n + \#_1(\omega)$ and in the case of Neumann boundary conditions on the $x$-axis $\nu = 2n + n$.

The relation between the code for the desymmetrized billiard and the symmetry classes is easily obtained using the symmetry lines. Therefore we first consider the relation between a symbol string and the symmetry lines $\mathcal{X}_0^>$, $\mathcal{X}_1^|$ and $\mathcal{T}_0$ in the desymmetrized system [20]. The line $\mathcal{X}_0^|$ corresponds to all points in $\tilde{\mathcal{P}}$, whose symbol sequence is symmetric with respect to the dot

$$\mathcal{X}_0^| \simeq \{\beta.\overleftarrow{\beta}\} \ . \tag{51}$$

The line $\mathcal{X}_1^>$ corresponds to all points in $\tilde{\mathcal{P}}$, whose symbol sequence is symmetric with respect to the letter 0

$$\mathcal{X}_1^> \simeq \{\beta.0\overleftarrow{\beta}\} \ . \tag{52}$$

The line $\mathcal{T}_0$ corresponds to all points in $\tilde{\mathcal{P}}$, whose symbol sequence is symmetric with respect to the letter 1

$$\mathcal{T}_0 \simeq \{\beta.1\overleftarrow{\beta}\} \ . \tag{53}$$

Iterating the basic symmetry lines $\mathcal{X}_0^>$, $\mathcal{X}_1^|$ and $\mathcal{T}_0$ shifts the point of symmetry to the left or right. Thus we obtain for the structure of the code words of periodic orbits with a certain symmetry the list shown in table 3.

Now the structure of the code words for a given symmetry in the full system can be determined from the structure of the code words for the desymmetrized system. The result is summarized in table 4. Notice that if one constructs code words with arbitrary $\alpha$ of the form given for $C_{\text{X}}$, $C_{\text{TX}}$ or $C_{\text{T}}$, it is possible to obtain a code word with higher symmetry. Furthermore it is possible to obtain a code word which is not primitive. The only symmetry class which is not incorporated into the above scheme using the symmetry lines are orbits from $C_{\text{TX}}$. The reason is that their code in the desymmetrized system does not have any of the symmetries, like orbits from $C_{\text{no}}$. The only difference between them is that orbits from $C_{\text{no}}$ have an even number of 1, whereas orbits from $C_{\text{TX}}$ have an odd number of 1's in the desymmetrized code. Thus for a given code length orbits from $C_{\text{TX}}$ in the desymmetrized system occur much more often than their number in table 2 for the full billiard might suggest. Actually, in the reduced system orbits from $C_{\text{TX}}$ are even more frequent than those from $C_{\text{no}}$.

| n | full | $l$ | $u_\gamma$ | symmetry | desymmetrized |
|---|---|---|---|---|---|
| 2 | AB | 5.196 | 1.433 | $C_{\text{full}}^{\vert\perp}$ | 1 |
| 3 | AAB | 6.585 | 2.027 | $C_{\text{X}}^{\vert>}$ | 011 |
| 4 | AAAB | 7.103 | 2.520 | $C_{\text{X}}^{\Vert}$ | 0011 |
| 4 | AABB | 9.238 | 2.853 | $C_{\text{full}}^{>\perp}$ | 01 |
| 5 | AAABB | 10.378 | 3.553 | $C_{\text{X}}^{\vert>}$ | 00101 |
| 5 | AABAB | 11.983 | 3.437 | $C_{\text{X}}^{\vert>}$ | 01111 |
| 6 | AAAABB | 10.945 | 4.129 | $C_{\text{X}}^{>>}$ | 000101 |
| 6 | AAABAB | 12.708 | 3.999 | $C_{\text{X}}^{\Vert}$ | 001111 |
| 6 | AAABBB | 11.836 | 4.531 | $C_{\text{full}}^{\vert\perp}$ | 001 |
| 6 | AABABB | 14.220 | 4.234 | $C_{\text{T}}$ | 010111 |
| 7 | AAAAABB | 11.263 | 4.608 | $C_{\text{X}}^{\vert>}$ | 0000101 |
| 7 | AAAABAB | 13.082 | 4.463 | $C_{\text{X}}^{\vert>}$ | 0001111 |
| 7 | AAAABBB | 12.559 | 5.405 | $C_{\text{X}}^{\vert>}$ | 0001001 |
| 7 | AAABAAB | 13.725 | 4.541 | $C_{\text{X}}^{\vert>}$ | 0011011 |
| 7 | AAABABB | 15.261 | 4.941 | $C_{\text{no}}$ | 0010111 |
| 7 | AABAABB | 16.003 | 4.931 | $C_{\text{X}}^{\vert>}$ | 0101011 |
| 7 | AABABAB | 17.224 | 4.835 | $C_{\text{X}}^{\vert>}$ | 0111111 |
| 8 | AAAAAABB | 11.457 | 5.015 | $C_{\text{X}}^{>>}$ | 00000101 |
| 8 | AAAAABAB | 13.297 | 4.851 | $C_{\text{X}}^{\Vert}$ | 00001111 |
| 8 | AAAAABBB | 12.958 | 6.196 | $C_{\text{X}}^{\Vert}$ | 00001001 |
| 8 | AAAABAAB | 14.017 | 4.973 | $C_{\text{X}}^{>>}$ | 00011011 |
| 8 | AAAABABB | 15.797 | 5.539 | $C_{\text{no}}$ | 00010111 |
| 8 | AAAABBBB | 13.347 | 6.687 | $C_{\text{full}}^{>\perp}$ | 0001 |
| 8 | AAABAABB | 16.848 | 5.559 | $C_{\text{no}}$ | 00101011 |
| 8 | AAABABAB | 18.000 | 5.420 | $C_{\text{X}}^{\Vert}$ | 00111111 |
| 8 | AAABABBB | 16.449 | 5.826 | $C_{\text{T}}$ | 00100111 |
| 8 | AAABBABB | 17.326 | 5.790 | $C_{\text{X}}^{\Vert}$ | 00101101 |
| 8 | AABAABAB | 18.588 | 5.457 | $C_{\text{X}}^{\Vert}$ | 01111011 |
| 8 | AABABABB | 19.367 | 5.619 | $C_{\text{T}}$ | 01011111 |
| 8 | AABABBAB | 19.105 | 5.576 | $C_{\text{full}}^{>\perp}$ | 0111 |

Table 5: Code words of periodic orbits and their length, stability exponent and symmetry class in the full system up to code length 8.

In table 5 we list all periodic orbits up to code length 8 together with their length, stability exponent and symmetry class of the full system and the corresponding code in the desymmetrized system.

## 4.2 Families of periodic orbits

Of special interest for the application of Gutzwiller's periodic orbit theory [1] are families of periodic orbits which accumulate in length. The code words of the simplest type of families have the form $\overline{A^n\omega}$, with $\omega$ starting and ending in $B$. One family corresponds to fixed $\omega$ and varying $n$. Some examples are shown in fig. 14. The family of orbits $\overline{A^nBB}$ yields the shortest accumulation length $l = 12$. Moreover this is the only family of the form $\overline{A^n\omega}$ with finite $\omega$, which exists for arbitrary $n$. All the other families of this kind are pruned for a finite $n > n_{\max}$.

This is nicely seen using the representation of the families in the symbol plane. One has

$$\gamma_n = \frac{N(A^n\omega)}{2^{n+m}-1} = \frac{1}{2^{n+m}-1}N(\omega) \tag{54}$$

$$\delta_n = \frac{N(A^n \overleftarrow{\omega})}{2^{n+m}-1} = \frac{2^n}{2^{n+m}-1}N(\overleftarrow{\omega}) \; , \tag{55}$$

and thus $\lim_{n\to\infty}(\delta_n, \gamma_n) = (\delta, 0)$, with $\delta = 2^{-m}N(\overleftarrow{\omega})$. Furthermore, it turns out, that all the points $(\delta_n, \gamma_n)$ lie on straight lines

$$\gamma_n = \frac{N(\omega)}{N(\overleftarrow{\omega})}\left(2^m\delta_n - N(\overleftarrow{\omega})\right) \; . \tag{56}$$

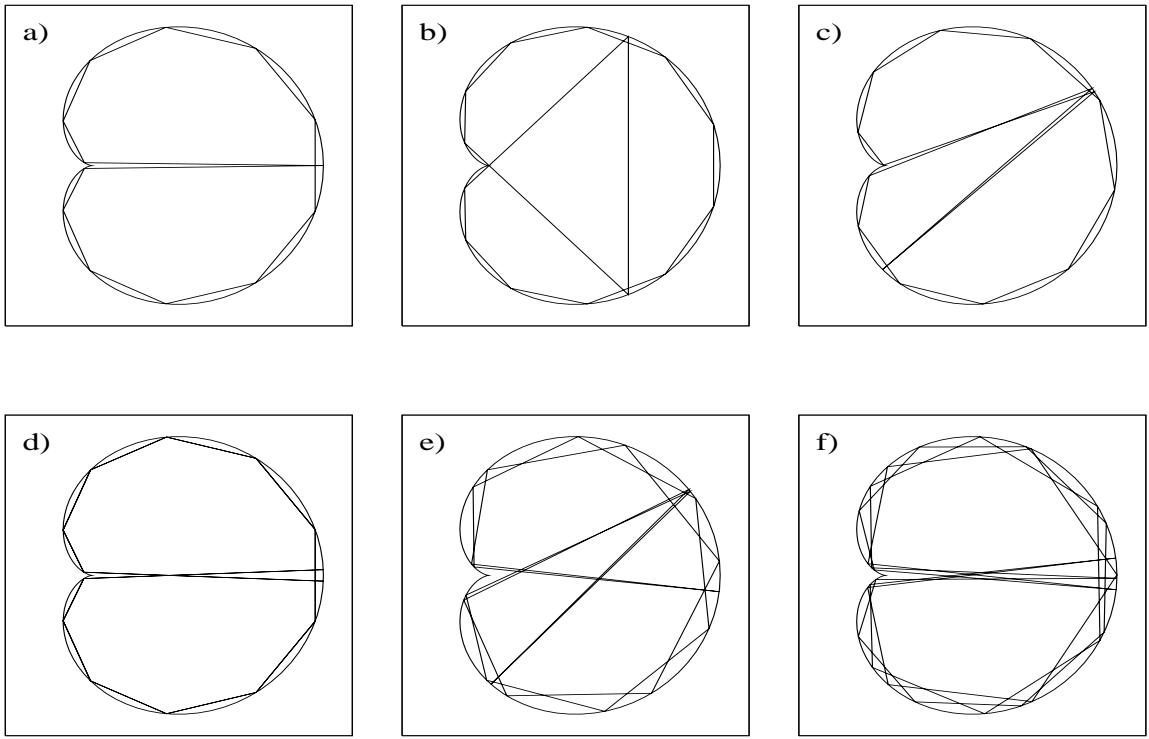

Figure 14: Examples of members of some families of periodic orbits which accumulate in length a) $\overline{A^nBB}$ for $n = 11$, b) $\overline{A^nBBB}$ for $n = 11$, c) $\overline{A^nBABB}$ for $n = 11$, d) $\overline{A^nBAB^mAB}$ for $n = m = 11$, e) $\overline{A^nBABAB^mAB}$ for $n = 7, m = 6$ f) $\overline{A^9B^2A^7BAB^6AB}$.

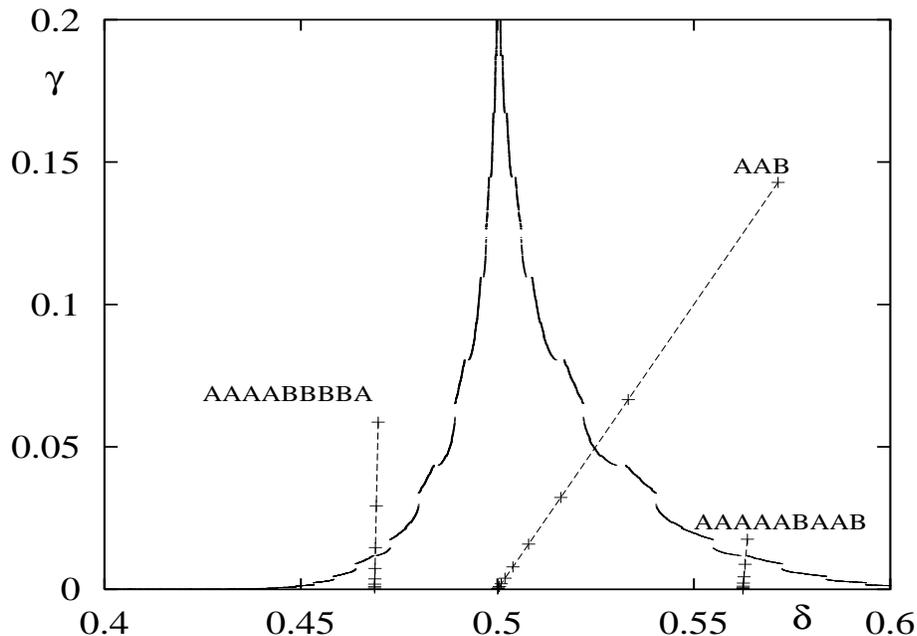

Figure 15: Sequence of points $(\delta_n, \gamma_n)$ in the symbol plane of some pruned accumulating families $\overline{A^n \omega}$ together with the pruning front. The labels denote the corresponding shifted code word of the first plotted member.

By shifting the code word or by considering a symmetric partner it is possible to obtain $\delta \in\; ]0.375, 0.75]$. If different words fulfill this criterion the one with the smallest distance to $\delta = 0.5$ is chosen.

In fig. 15 some examples are shown together with the pruning front. From the figure one can read off the last allowed orbit as well as the pruning mechanism. If $\delta \in ]0.375, 0.5[$ the family is s–pruned, whereas for $\delta \in ]0.5, 0.75[$ the family is f–pruned. As explained above it is also possible that a family is pruned by both mechanisms.

We observe that the pruning front is a monotonously increasing curve in the interval $[0.375, 0.5]$, and monotonously decreasing for $[0.5, 0.75]$. Moreover it only reaches 0 at 0.375 and 0.75. Therefore families of *any* kind can only exist if at least one of their limit points in the symbol plane is $(0.75, 0)$. In particular we think that the family $\overline{A^n BB}$ exists for any $n$.

One can construct a variety of more complicated families, e.g. with the structure $\overline{A^n \alpha A^m \beta}$ or $\overline{A^n \alpha B^m \beta}$, see fig. 14. The special case $\overline{A^n BAB^n AB}$, see fig. 14e), has the accumulating length $l = 24$ and probably exists for any $n$, because the corresponding limit point in the symbol plane is $(0.75, 0)$. Actually we think that there is an infinite number of accumulating families. They can be constructed by introducing two combined symbols $F_n = AB^n A$ and $G_m = BA^m B$. Now *any* sequence of $F_n$'s and $G_m$'s gives an accumulating family, as long as all the indices collectively go to infinity. One example with different $n_i$ is shown in fig. 14f). The length of the limit orbit of such a family is given by the total number of the symbols $F_n$ and $G_m$ multiplied by 12.

The stability of periodic orbits from accumulating families grows with increasing period, because they must come arbitrary close to the cusp. However, their stability grows much slower than for generic orbits because they also get arbitrary close to the parabolic line $\mathcal{F}$.

The pruning of the "almost accumulating" families is related to the limiting orbit which almost hits the cusp. It consists of a sliding motion and of a finite orbit hitting the cusp as illustrated in fig. 14. We now turn to the study of these finite orbits.

## 4.3 Finite orbits

In the quantum mechanical billiard problem discontinuities of derivatives of the boundary curve can play an important role (see, e.g. [25, 26, 27] and [28, 29, 30, 12, 31] and references therein). The cusp orbits start at an arbitrary angle in the cusp and eventually return to the cusp with an arbitrary angle without the need to fulfill any reflection condition in the cusp. We think that the investigation of orbits having a reflection in the cusp is already important from the classical point of view: on the one hand due to the connection to families of periodic orbits and on the other hand because cusp orbits are also extrema of the Lagrangian, as it is the case for periodic orbits. This can be seen in the following way: the variation of the Lagrangian $\mathcal{L} = \mathcal{L}(\phi_1, \ldots, \phi_n)$ yields a system of $n$ coupled nonlinear equations $\nabla \mathcal{L} = 0$. If $\phi_i = \pm\pi$ the corresponding equation is automatically fulfilled,

$$\left.\frac{\partial \mathcal{L}}{\partial \phi_i}\right|_{\phi_i=\pm\pi} = \left.\frac{\partial \tau(\phi_{i-1}, \phi_i)}{\partial \phi_i} + \frac{\partial \tau(\phi_i, \phi_{i+1})}{\partial \phi_i}\right|_{\phi_i=\pm\pi} = 0 \qquad (57)$$

regardless of the values of $\phi_{i-1}$ and $\phi_{i+1}$. This means that there is no reflection condition to fulfill in the point $\phi_i = \pm\pi$. Thus in some sense the singularity looks like an infinitely small circle, i.e. it can reflect into any direction. Notice that the above argument only holds if we look at $\mathcal{L}$ as a function of $\phi_i$. This dependency of the Lagrangian on the choice of coordinates is due to the singularity. We will not consider cusp orbits as periodic orbits, but instead as finite orbits with the same initial and final point in configuration space.

In order to classify the cusp orbits we extend the symbolic dynamics with a third letter $C$, which corresponds to the lines $\mathcal{S}$ and $\Gamma$. For a cusp orbit starting in $\mathcal{S}$ the next point lies on $\Gamma^{-1}$, from which one iterates until the first point lies on $\Gamma$. Since the momentum is not defined in the next step we consider cusp orbits as finite orbits whose code word is $C\alpha C$, where $\alpha$ is a code word consisting of $A$ and $B$. One could omit the $C$'s, but we find it more convenient to have the number of letters in a code word correspond to the number of reflections. Thus the shortest cusp orbit, which runs along the symmetry line has the code $CC$, because it starts in $\mathcal{S}$ and has its last reflection on $\Gamma$, see fig. 16a).

Since the partition of the Poincaré section as shown in fig. 6 and fig. 7 is given by the iterates of $\Gamma$, cusp orbits correspond to the intersections of these lines. Actually it is simplest just to iterate $\Gamma^{-1}$ and to look of intersections with $\Gamma$. After $n$ iterations we have $2^{n+1} - 1$ intersections (ignoring pruning) i.e. at every iteration there are $2^n$ new intersections, which is exactly the number of cusp orbits one expects without pruning. Defining $\Gamma^{-n-1} = P^n \Gamma^{-1}$ the number of reflections of a cusp orbit which is given by the intersection of $\Gamma^{-n-1}$ and $\Gamma$, is $n + 2$.

Similar to the case of periodic orbits one can classify finite orbits hitting the cusp according to their symmetry by using the symmetry lines. Intersections of $\mathcal{T}_0$ and $\Gamma^{-n}$ correspond to $T$-symmetric cusp orbits which have period $2n$ because the distance to $\Gamma$ and $\Gamma^{-1}$ is $n-1$ forward respectively backward iterations. Intersections of $\mathcal{X}_0$ and $\Gamma^{-n}$ yield $X$-symmetric cusp orbits. In the same way cusp orbits with $\mathcal{X}_1$ symmetry are found. If one starts on a symmetry line past and future are symmetric and thus the symmetry is reproduced in the cusp, i.e. a $T$-symmetric orbit retraces itself also in the cusp and a $X$-symmetric orbit has a $X$-symmetric reflection in the cusp. Note that the latter use the vertical tangent in $\mathcal{S}$, while the others use any tangent vector. The exceptional period two cusp orbit $CC$ has full symmetry. This is not the case for any other cusp orbit.

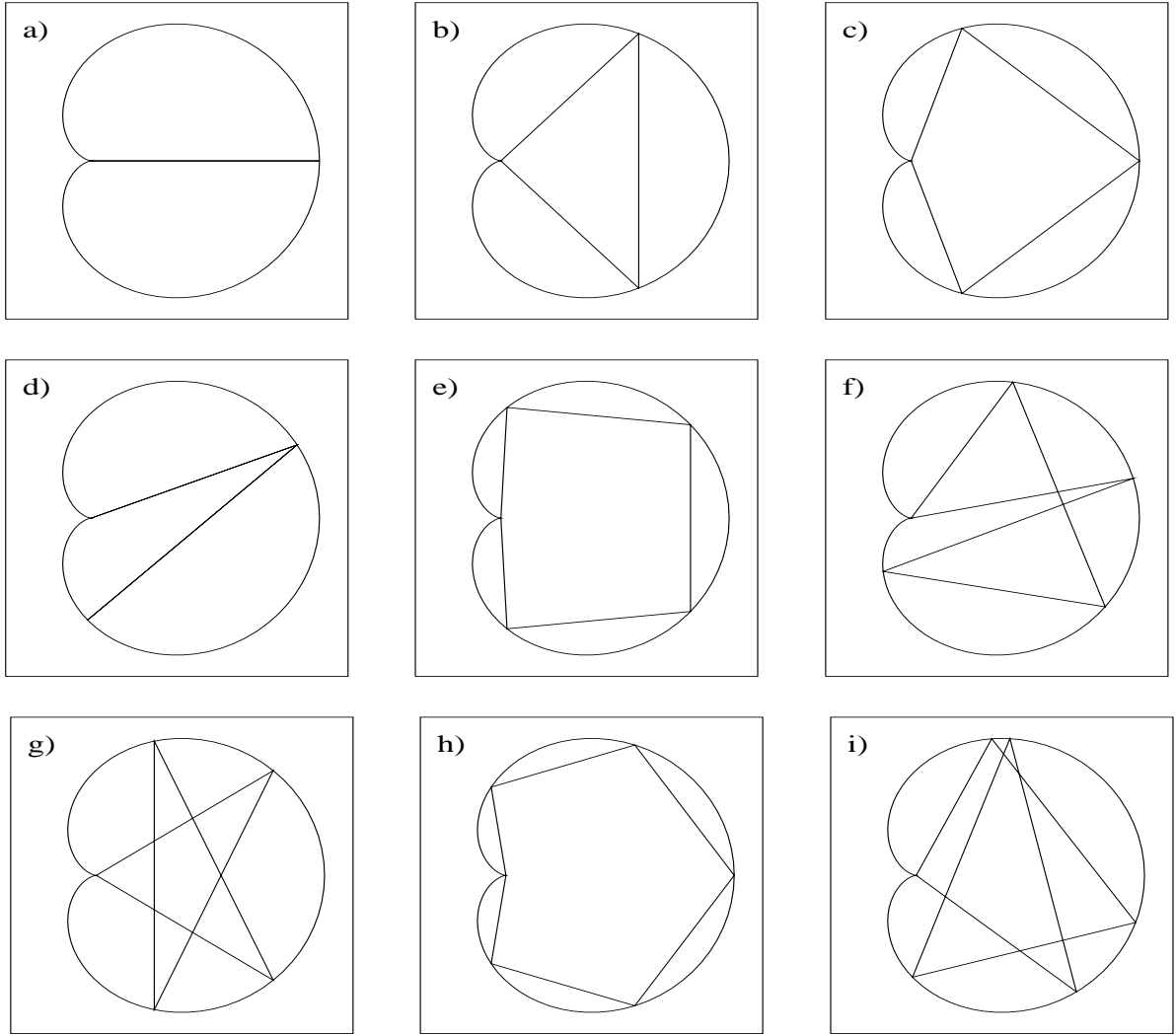

Figure 16: Examples of finite orbits a) $CC$, b) $CAC$, c) $CAAC$, d) $CABC$, e) $CAAC$, f) $CAABC$, g) $CABAC$, h) $CAAAAC$ and i) $CAABAC$.

By counting the number of intersections of the symmetry lines and iterates of $\Gamma$ (ignoring pruning) we find that there are $2^{(n-2)/2}$ symmetric cusp orbits of even period $n$ and $2^{(n-3)/2}$ for odd $n$. Like in the case of periodic orbits the non-symmetric cusp orbits are the overwhelming majority for large $n$. For a complete symbolic dynamics the number of cusp orbits is exactly $2^{n-2}$, where $n$ is the total length of the word $C\alpha C$. So already this number grows faster then the number of periodic orbits because we do not have to consider cyclic classes here. Furthermore cusp orbits can be combined to give multiple cusp orbits. Their number can be estimated using the topological polynomial [32] for the symbolic dynamics with letters $\{A, B, G = CC\}$, which is $1 - 2z - z^2$, such that we obtain $h_{\text{top}} = \ln(1 + \sqrt{2})$ for the topological entropy. Thus they occur much more often than the periodic orbits for which the $h_{\text{top}} = \ln 2$ in the case of a complete symbolic dynamics. Actually we do not gain any information by this because every multiple cusp orbit can be uniquely decomposed into single cusp orbit, but classically they become interesting if we smoothen the singularity and quantum mechanically they are interesting as multiple diffractive orbits. Finally note that in considering multiple cusp orbits one can produce full symmetry and TX-symmetric cusp orbit by superposition of single cusp orbits with their appropriate symmetric partners. In this way we can even build multiple cusp orbits with any symmetry starting with a cusp orbit with no symmetry.

We will now briefly explain the relation between cusp orbits and the families of accumulating orbits. The families $\overline{A^n \omega}$ perform a large number of successive reflections along the boundary, which are followed by some other reflections after which the orbit returns to the approximate sliding motion. For large $n$ this family tends to the sliding motion plus a limiting cusp orbit. If $\overline{A^n B \alpha B}$ is the code word for the family, the corresponding cusp orbit is given by $C \alpha C$. The largest angle $\chi$ between the $x$-axis and the limiting cusp orbit determines the maximum number of consecutive reflections having the same letter. Only for $\chi = 0$ a family can exist for any $n$. Otherwise we have a finite family of short orbits which do not accumulate.

# 5  Global properties of the billiard

In this section some global characteristics of the cardioid billiard are calculated. In the first part we obtain an analytical result for the average distance between reflections $\bar{l}$ and an analytical estimate for the Kolmogorov-Sinai entropy $h_{\text{KS}}$. The same quantities are then calculated using averages over periodic orbits, where each orbit is weighted by its stability.

We first calculate dynamical averages by iterating the billiard map. Averaging over several generic trajectories for $5 \times 10^6$ iterations we obtain for the average length between reflections $\bar{l}$ and the Lyapunov exponent, which in our case approximates the KS entropy $h_{\text{KS}}$

$$\bar{l} \approx 1.851 \tag{58}$$
$$h_{\text{KS}} \approx 0.653. \tag{59}$$

$h_{\text{KS}}$ is calculated according to [33], however, we use the linearized map (16). One can also calculate these quantities analytically:

For any billiard in the domain $\Omega$ the average length is determined by the area $|\Omega|$ and circumference $|\partial \Omega|$ by transforming the average over the length function $l(s,p)$ on $\mathcal{P}$ into the volume of the energy shell [33]

$$\bar{l} = \int_{\mathcal{P}} l(s,p) \mathrm{d}\mu = \frac{1}{2|\partial \Omega|} \int_{\mathcal{P}} l(s,p) \, \mathrm{d}s \, \mathrm{d}p \tag{60}$$

$$= \frac{1}{2|\partial \Omega|} \int_{\Omega \times S^1} \mathrm{d}x \, \mathrm{d}y \, \mathrm{d}\beta = \frac{|\Omega|\pi}{|\partial \Omega|}, \qquad \text{i.e. for the cardioid} \tag{61}$$

$$= \frac{3}{16}\pi^2 = 1.85055... \tag{62}$$

where $\mathrm{d}\mu = \frac{1}{2|\partial \Omega|} \mathrm{d}s \mathrm{d}p$ is the normalized Liouville measure on $\mathcal{P}$.

An analytical expression for a lower bound on $h_{\text{KS}}$ was given by Wojtkowski in [3]:

$$h_{\text{KS}} \geq \int_{\mathcal{P}} \log \varrho \, \mathrm{d}\mu \;, \tag{63}$$

where $\varrho$ estimates the contribution to the expansion from below. Since $\varrho$ contains the length $l(s,p)$, which is given by the root of a cubic equation we again pass to the generating function of the map $l(\phi_1, \phi_2)$. The variables are transformed by

$$s(\phi_1, \phi_2) = 4\sin(\phi_1/2) \tag{64}$$

$$p(\phi_1, \phi_2) = \langle v, T_1 \rangle = \frac{\langle L(\phi_1, \phi_2), T(\phi_1) \rangle}{l(\phi_1, \phi_2)}. \tag{65}$$

Since $s$ is independent of $\phi_2$ we obtain for the Jacobian of the transformation

$$\det \frac{\partial(s,p)}{\partial(\phi_1, \phi_2)} = 2\cos(\phi_1/2) \left( \langle L', T \rangle - \frac{\langle L, T \rangle \langle L, L' \rangle}{l^2} \right) / l, \tag{66}$$

where a prime denotes the derivative with respect to $\phi_2$. Since not all combinations $(\phi_1, \phi_2)$ correspond to allowed trajectories, we have to restrict the integration ranges. Moreover we use the symmetry of the system to restrict the integration to positive $\phi_1$ and obtain

$$h_{\mathrm{KS}} \geq \frac{1}{|\partial \Omega|} \int_0^\pi \int_{-\pi+\phi_1}^\pi \varrho(\phi_1, \phi_2) \det \frac{\partial(s,p)}{\partial(\phi_1, \phi_2)} \mathrm{d}\phi_2 \mathrm{d}\phi_1. \tag{67}$$

Let the linearized Poincaré map be given by

$$DP = \begin{pmatrix} a & b \\ c & d \end{pmatrix} \tag{68}$$

with $\det DP = 1$. The arguments of Wojtkowski [3] for matrices with only positive (or negative) entries work equally well for checker board matrices (18). Then $\varrho = \sqrt{ad} + \sqrt{bc}$. Using (16) we obtain $ad = (k_1 - 1)(k_2 - 1)$ where $k_i = t\kappa_i/n_i$. Therefore we have to evaluate the following integral

$$h_{\mathrm{KS}} \geq \int_0^\pi \int_{-\pi+\phi_1}^\pi \log\left( \sqrt{(k_1-1)(k_2-1)} + \sqrt{k_1 k_2 - k_1 - k_2} \right) \det \frac{\partial(s,p)}{\partial(\phi_1, \phi_2)} \mathrm{d}\phi_2 \mathrm{d}\phi_1. \tag{69}$$

The numerical integration gives $h_{\mathrm{KS}} \geq 0.633$. Even though Wojtkowski's theorem only gives a lower bound on $h_{\mathrm{KS}}$ this value is surprisingly close to the numerically measured value.

$h_{\mathrm{KS}}$ can also be approximated by calculating the entropy of the partition of the Poincaré section $\mathcal{P}$ into cells labeled by words of length $n$

$$h_{\mathrm{KS}}(n) = -\frac{1}{n} \sum_\omega p_\omega \ln p_\omega, \tag{70}$$

where $p_\omega = \mu(\omega)$ is the size of the cell with label $\omega$ [34]. Since the cell sizes in $\mathcal{P}$ are quite hard to obtain we determine them numerically in the symbol plane. This is done by calculating the probability with which each cell is visited by performing many iterations of one initial condition. Thus we numerically approximate the invariant measure in the symbol plane by assigning a probability $p_\omega$ to every cell labeled by a word $\omega$ of length $n$. Thus the sum (70) extends over all allowed cells in the symbol plane. If we denote the total number of allowed cells labeled by words of length $n$ by $N(n) = \sum_\omega 1$, we obtain an estimate for the topological entropy with respect to code length with $p = 1/N(n)$

$$h_{\text{top}}(n) = -\frac{1}{n}\sum_\omega p \ln p = \frac{1}{n} \ln N(n) \;, \tag{71}$$

For $n = 22$ we obtain $h_{\text{KS}} \approx 0.66$ and $h_{\text{top}} \approx \ln(1.98) \approx 0.68$. $h_{\text{KS}}$ is smaller than $h_{\text{top}}$; the difference tells us how much the invariant measure on the symbol plane deviates from equipartition. The three values for $h_{\text{KS}}$ are $0.63 < 0.65 < 0.66$. Their ordering is correct because the analytical value is a lower bound and the value obtained from the symbol plane tends to overestimate due to finite size effects.

## 5.1 Global properties from periodic orbits

Now we repeat the calculation of $h_{\text{KS}}$, $h_{\text{top}}$ and $\bar{l}$ using periodic orbits. The main point in the calculation of these averages is that the sum in (70) can be transformed into a sum over periodic orbits of fixed length. Each cell visited by a periodic orbit $\gamma$ is assigned a probability given by the inverse of the eigenvalues $e^{-u_\gamma}$, from which the invariant measure is approximated (see, e.g., [35, 32, 36]).

This approach can be illustrated by a plot of the distribution of periodic orbits in the Poincaré section (see fig. 17). We observe that the distribution is not as uniform as one might expect. Most obvious we have regions with very few periodic orbits in the neighborhood of $\mathcal{F}$, because these orbits are rather stable. The same structure can be observed in the division of $\mathcal{P}$ into cells in fig. 7. In a region where the cells are small there is a high density in contrast to regions with large cells, where the density is low. As we already discussed the accumulating families are not only close to $\mathcal{F}$ but also close to $\mathcal{S}$ and therefore close to $\Gamma$ and its images and preimages. These families are the only ones that evade the region surrounding $\Gamma$ exactly at the points where there exists a corresponding cusp orbit. Parts of $\Gamma$ with large $s$ have a high density of periodic orbits in their neighborhood. They come close to $\Gamma$ without having been close to $\mathcal{F}$, and are therefore extremely unstable, which is the reason for their high density. Moreover some of the basic symmetry lines and their iterates are visible. The reason for a relatively high density of points along the symmetry lines is the fact that they are one-dimensional. Thus the small fraction of symmetric orbits has to fill only a "small" set in $\mathcal{P}$, whereas the large number of orbits without symmetry has to fill large areas in $\mathcal{P}$. Plotting only orbits of higher period the symmetry lines are less visible. If one plots the periodic orbits in the symbol plane, they are much more uniformly distributed, see fig. 18.

The weighted average over periodic orbits of length $n$ of a quantity quantity $f(u, l)$ is now given by

$$\langle f \rangle_n^w = \frac{\sum_\gamma f(u_\gamma, l_\gamma) e^{-u_\gamma}}{\sum_\gamma e^{-u_\gamma}} \;, \tag{72}$$

where the sum runs over all periodic orbits $\gamma$ of code length $n$, including multiple traversals. Although for large $n$ we do have $\sum_\gamma e^{-u_\gamma} \approx 1$ [37] we include this term for normalization.

Using the periodic orbits up to code length 20 we we calculated the mean stability exponent per reflection $\langle u \rangle_n^w / n$. The values approach a constant for increasing $n$ with $\langle u \rangle_{20}^w / 20 = 0.657$, which agrees quite well with the the value of $h_{\text{KS}}$ in (59).

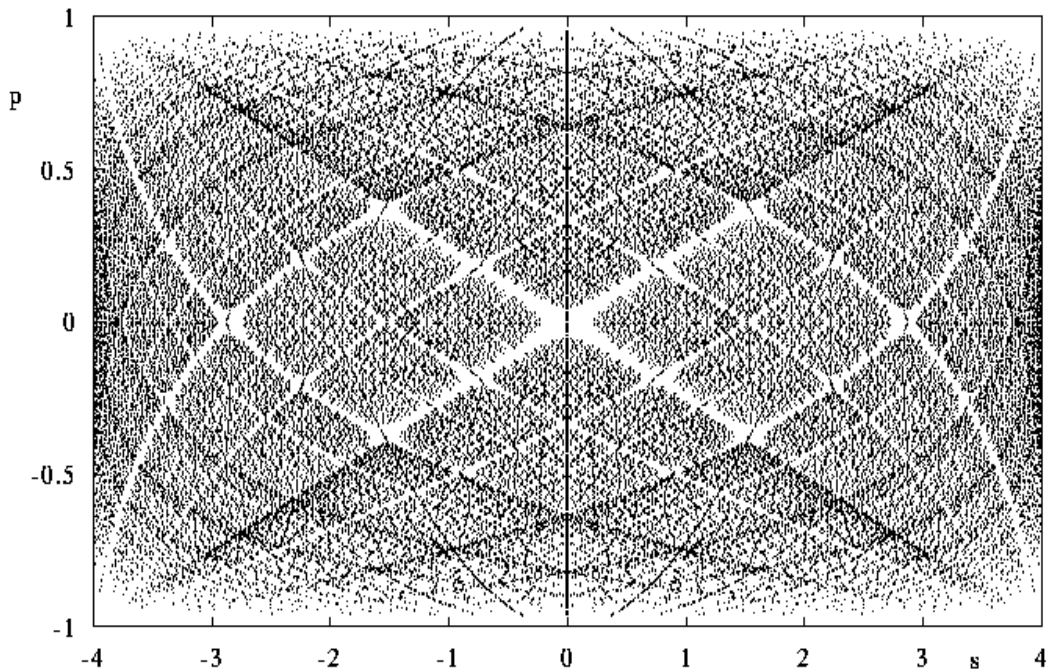

Figure 17: Plot of the points of all periodic orbits with code length up to $n = 15$ in the Poincaré section $\mathcal{P}$. The symmetry lines and the $\Gamma^n$ are visible, comp. to fig. 4.

For the average distance between consecutive reflections we obtain $\langle l \rangle_{20}^w / 20 = 1.864$, where again the values are slowly decreasing for increasing $n$, such that we have good agreement with the analytical result (62). In fact the conversion of values measured with respect to the discrete time (Poincaré section) and the one measured in phase space is just given by the factor (62). Thus the corresponding quantities for the continuous flow are given by

$$h_{\text{KS}}(20)/\bar{l} = 0.36 \tag{73}$$

$$\frac{1}{20}\langle u \rangle_{20}^w / \bar{l} = 0.36. \tag{74}$$

Of course $h_{\text{KS}}/\bar{l}$ can also be approximated by averaging $u/l$ which gives $\langle u/l \rangle_{20}^w = 0.363$.

For the family of billiards (1) Robnik calculated the KS-entropy in [7]. He gave values for $\epsilon$ close to 1, from which we extrapolate $h_{\text{KS}} = 0.34$ for the cardioid, which is consistent with our values.

Finally we calculate the topological entropy $h_{\text{top}}$ from the growth of the number of periodic orbits by considering $h_{\text{top}}(n) = \frac{1}{n} \ln N(n)$, where $N(n)$ is the total number of points belonging to all the periodic orbits of period $n$, including multiple traversals. We obtain a plateau for $n \geq 10$ with $h_{\text{top}} \approx 0.683 \approx \ln 1.98$, which is in good agreement with the value obtained in the symbol plane.

## 5.2 Statistical properties of periodic orbits

The following investigations are based on periodic orbits up to code length 20 (See, e.g. [14, 23, 38, 39, 40] for similar studies for other systems). The principle difference of the averages $\langle\rangle^o$ taken in this section to the averages $\langle\rangle^w$ calculated in the previous section is that now we take all periodic orbits as equally weighted, i.e. we have a uniform weight in the symbol plane. Thus we transfer the picture of periodic orbits in the Poincaré section (fig. 17) into the symbol plane (fig. 18) and average $f(u,l)$ over periodic orbits $\gamma$ of code length $n$ by

$$\langle f\rangle_n^o = \frac{1}{N(n)} \sum_\gamma f(u_\gamma, l_\gamma). \tag{75}$$

Looking at fig. 18 we first observe that in fact the orbits are distributed rather uniformly in the non pruned region. The symmetry lines are almost invisible. Only the accumulating families show up as holes that are not related to the pruning front.

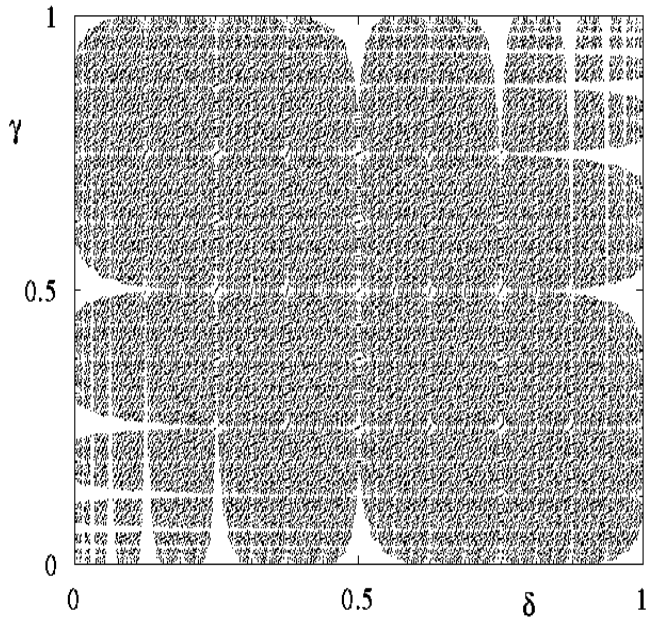

Figure 18: Plot of the points of all periodic orbits up to code length $n = 15$ in the symbol plane, compare to fig. 17 where the corresponding points are presented in the Poincaré section.

In the last section we calculated the growth behaviour in terms of the code length, but also the growth behaviour of the number $\mathcal{N}(l)$ of periodic orbits $\gamma$ with geometric length $l_\gamma$ below a given length $l$ is of interest. Because the cardioid billiard is strongly chaotic one expects the typical exponential proliferation

$$\mathcal{N}(l) \sim \mathrm{Ei}\,(\tau l) \sim \frac{e^{\tau l}}{\tau l}\;, \tag{76}$$

where $\tau$ is the topological entropy with respect to the geometric length of orbits. However, this spectral staircase is not well defined in our case because of the families accumulating in length (the same happens, e.g. in the case of the stadium billiard or the wedge billiard [39]).

Therefore we define $\Xi_{\mathrm{acc}}$ as the set of all periodic orbits $\gamma$ whose corresponding code has more than five consecutive letters $A$ or $B$, and $\Xi_{\mathrm{reg}}$ as the set of periodic orbits with up to five consecutive $A$ or $B$. This choice is somewhat arbitrary, but motivated by the observation that orbits from $\Xi_{\text{acc}}$ already have the geometrical structure in configuration space of the possible limit orbit of the considered family. We define $\mathcal{N}_{\text{reg}}(l)$ as the counting function of the number of orbits from $\Xi_{\text{reg}}$ with geometric length less than $l$. In fig. 19 $\mathcal{N}(l)$ and $\mathcal{N}_{\text{reg}}(l)$ are shown in logarithmic representation using the periodic orbits up to code length 20 together with a fit with the asymptotic behaviour (76) $\tau \approx 0.345$ for $\mathcal{N}_{\text{reg}}(l)$. One clearly observes for $\mathcal{N}(l)$ the strong increase next to length $l = 12$, which is caused by the family $\overline{A^n BB}$. This step would even be more pronounced, if orbits of higher code length had been used. In contrast the logarithmic plot of $\mathcal{N}_{\text{reg}}(l)$ is a much "smoother" curve, with most of these "steps" removed.

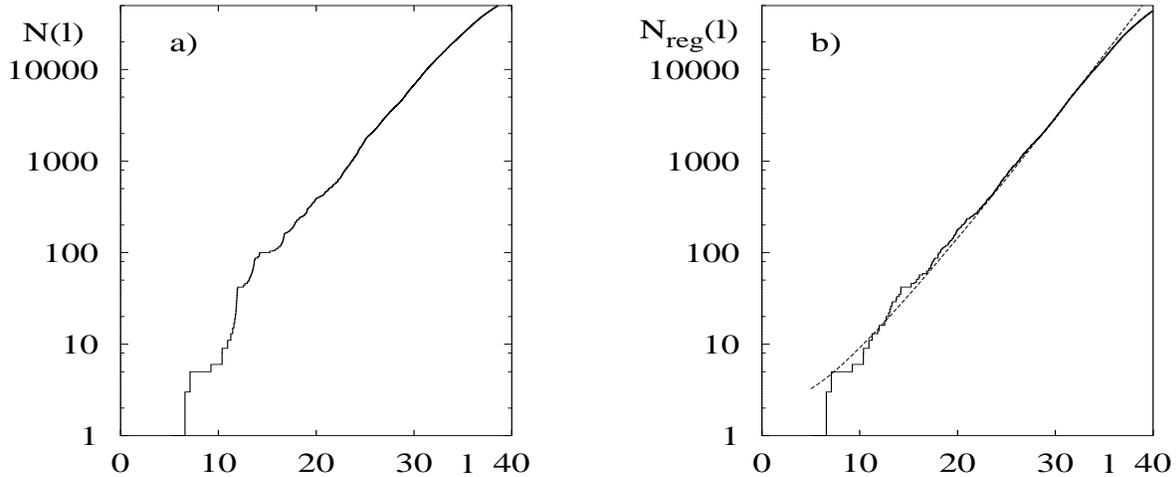

Figure 19: In a) the logarithmic representation of the number of periodic orbits with geometric length less than $l$ is shown using all orbits up to code length $n = 20$. In b) $\mathcal{N}_{\text{reg}}(l)$ is shown in logarithmic representation where all orbits with more than five consecutive $A$ or $B$ are excluded; also shown is $\frac{e^{\tau l}}{\tau l}$ for $\tau = 0.345$.

That the above splitting allows one to select orbits belonging to accumulating families is also visible if one plots the distribution $p_n(l)$ of lengths for a given code length. In fig. 20a), where all orbits with code length 14 were used, long tails down to $l = 10$ are visible. In contrast in fig. 20b) only orbits from $\Xi_{\text{reg}}$ were used, and the outliers are removed.

We calculated the average length of periodic orbits and observed the expected increase of the mean length $\langle l \rangle_n^o = n\tilde{l}$ with the code length $n$, with $\tilde{l} = 1.99$ using all orbits and $\tilde{l} = 2.054$ using orbits from $\Xi_{\text{reg}}$. The variance of the distribution increases with $n$. For orbits from $\Xi_{\text{reg}}$ one finds that $\sigma_n^2 \approx \tilde{\sigma}^2 n$ for $n \geq 15$, with $\tilde{\sigma}^2 \approx 0.47$. Furthermore we observed, as in the case of the hyperbola billiard [14], that $p_n(l)$ are approximately Gaussian distributed. This is demonstrated in fig. 20d) where each $p_n(l)$ is shifted to the origin and the variance is normalized.

In [14] a relation between the number of primitive cyclic classes of the code words and the classical staircase was derived under the assumption that $p_n(l)$ has a Gaussian distribution and that the mean length $\langle l \rangle_n^o$ and the variance $\sigma_n^2$ depend linear on $n$. The result adopted to our case is

$$\tau \approx \frac{1}{\tilde{\sigma}^2}\left(\tilde{l} - \sqrt{\tilde{l}^2 - 2\tilde{\sigma}^2 h_{\text{top}}}\right) = \frac{h_{\text{top}}}{\tilde{l}} \frac{2}{1 + \sqrt{1 - 2h_{\text{top}}\tilde{\sigma}^2/\tilde{l}^2}} \quad , \tag{77}$$

where $h_{\text{top}}$ denotes the topological entropy with respect to code words. In the case of orbits

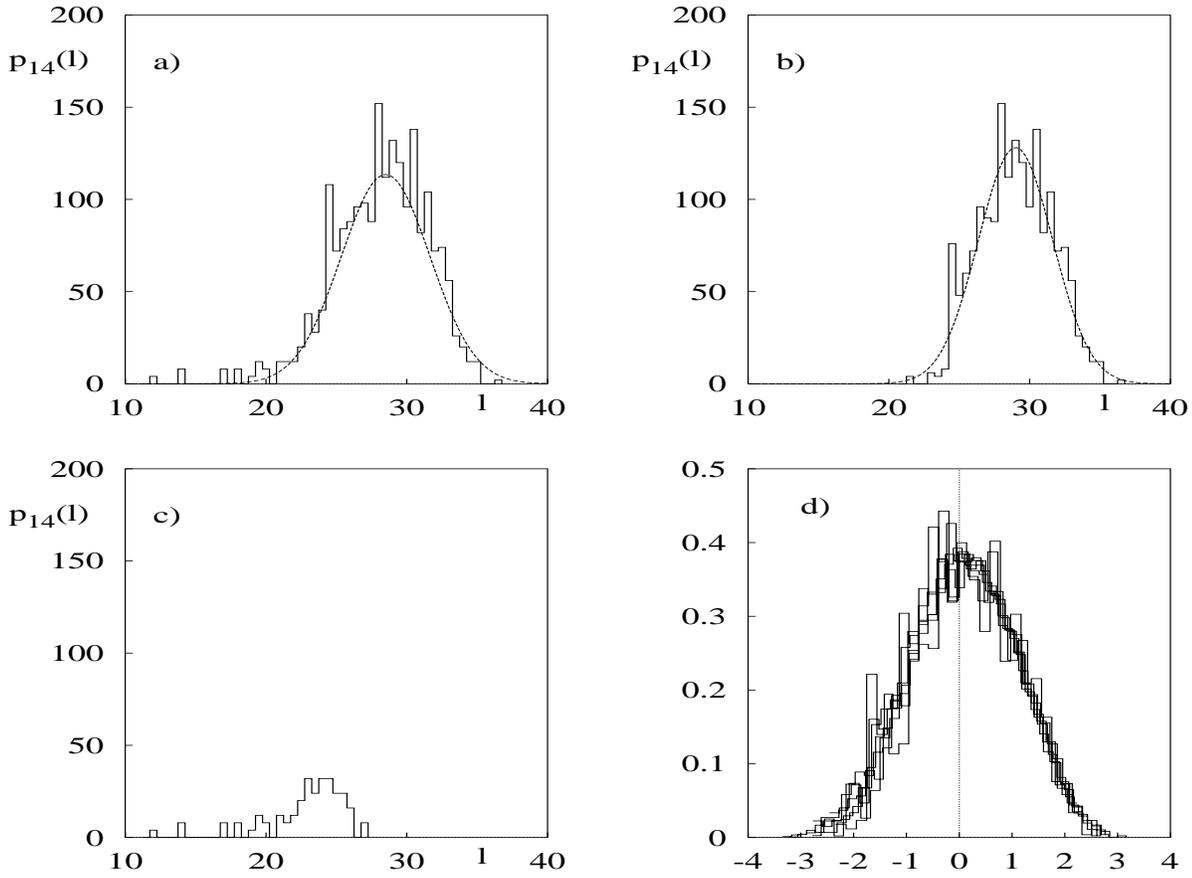

Figure 20: Probability distribution $p_n(l)$ for $n = 14$: a) all orbits are included, b) only orbits with not more than five consecutive $A$ or $B$ are used, and in c) the difference between both is shown. In addition in a) and b) a Gaussian distribution is shown as dashed line. In d) the shifted and normalized distributions $p_n(l)$ are shown for $n = 14, \ldots, 20$.

from $\Xi_{\text{reg}}$ we obtain $h_{\text{top}} = 0.672$ and therefore

$$\tau \approx 0.34 \ , \qquad (78)$$

which is in good agreement with the value $\tau \approx 0.35$ obtained from the fit of $\mathcal{N}_{\text{reg}}(l)$ with the asymptotic behaviour. The consistency of these results can be seen as justification of the above splitting.

We also considered the distribution of the stability exponents $u_\gamma$ for fixed code length $n$ and found that $u_\gamma$ is centered around a mean $\langle u \rangle_n^o = n\tilde{u}$ with $\tilde{u} = 0.70$ using either all orbits or only orbits from $\Xi_{\text{reg}}$. The distributions are stronger peaked than Gaussian distributions.

A further common statistics is the spacing between neighbouring lengths. In order to obtain a mean spacing of one, the asymptotic behaviour is used to unfold the length spectrum. The result using orbits from $\Xi_{\text{reg}}$ with length $l < 30$ is shown in fig. 21 is in agreement with the Poisson distribution $P(s) = e^{-s}$.

It is also interesting to look at the dependence of the stability exponents $u_\gamma$ vs. the geometric length $l_\gamma$. The result is shown in fig. 22. Some of the families are clearly visible as series of points with accumulating $l_\gamma$ and thereby increasing $u_\gamma$.

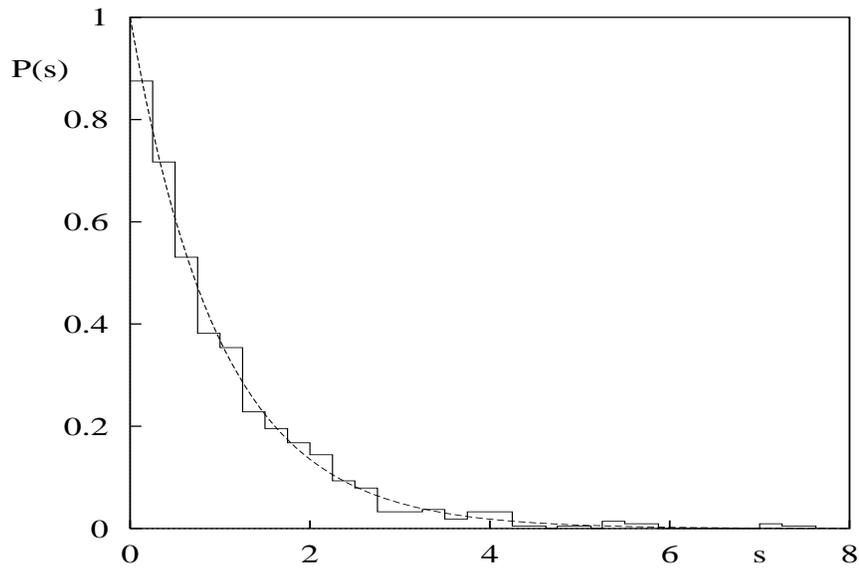

Figure 21: Length spacing distribution $P(s)$ (full line) together with the Poissonian distribution $e^{-s}$ (dashed line).

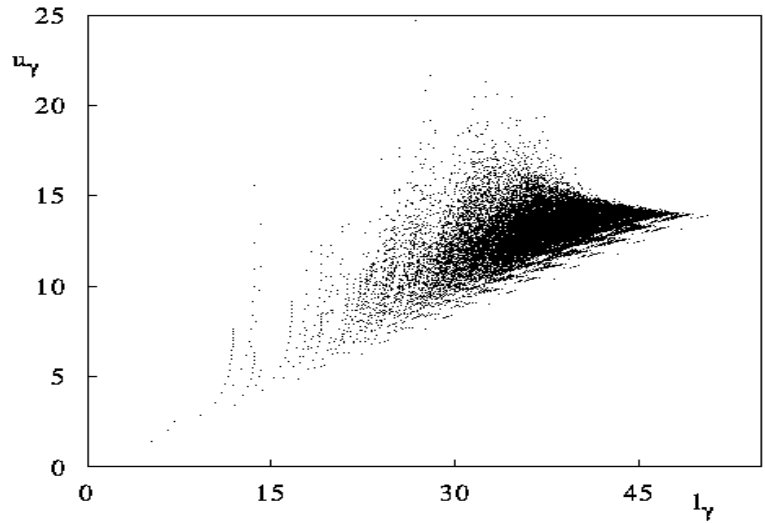

Figure 22: Plot of the stability exponents $u_\gamma$ vs. the geometric length $l_\gamma$ using all orbits up to code length 20.

# 6 Summary

In this paper we have developed a symbolic dynamics for the cardioid billiard by constructing a partition of the Poincaré section $\mathcal{P}$ using two symbols $A$ and $B$. After completion of the part on the symbolic dynamics a preprint appeared [12], where the same coding was found independently and used to study diffraction effects of the quantum mechanical system.

For the symbolic dynamics it turned out that not every sequence of symbols is allowed. We obtained the pruning front, which comes in two pieces related to the two pruning mechanisms in the system. Using the symbolic dynamics periodic orbits can be labeled in a unique way. Assisted by the knowledge that all orbits correspond to maxima of the Lagrangian we calculated a large number of periodic orbits up to code length 20. Complete sets of higher periodic orbits are hard to obtain because some of them are extremely unstable due to the unbounded curvature. Using the symmetry lines of the billiard map, we presented a classification of periodic orbits with respect to their symmetry properties. This was first obtained in the symmetry reduced symbolic dynamics and then translated back to the full system. Combining Wojtkowski's result about convex scattering billiards with a geometric argument in the symmetry reduced system we were able to determine the Maslov indices from the code words.

Studying families of periodic orbits with short geometric length provides a nice application of the pruning front, because it allowed us to determine whether a family exists for arbitrary code length, or whether it is eventually pruned. In the latter case a plot of the points of the periodic orbits in the symbol plane allows for the determination of the last allowed member of the family. The converse argument enables us to write down an infinite number of families that accumulate in length. Furthermore it turned out that finite respectively cusp orbits appear rather natural even from a merely classical point of view as parts of the possible limit orbits of the short families.

In the last section we calculated an estimate for the Kolmogorov-Sinai entropy and found good agreement with numerically calculated values. Averaging the periodic orbits we find consistent values for the KS entropy and the average length between reflections. We illustrate the idea of the periodic orbit averaging by a plot of periodic orbits in the surface of section. The topological entropy is quite close to $\ln 2$ because the pruning sets in rather late.

To obtain well defined statistics despite the presence of accumulating families we suggested a method to subtract the accumulating families. This procedure might also be helpful when using Gutzwiller's periodic orbit theory [1], where families of accumulating periodic orbits have to be treated separately. We believe that this method might be useful for other systems with accumulating families.

With the complete set of periodic orbits, knowledge of the Maslov indices and an understanding of accumulating families and cusp orbits all the ingredients for the periodic orbit quantization of the cardioid billiard are available. Eventually the same program should be carried out for Robnik's family of billiards, at least close to the cardioid, where it is conjectured to be ergodic. On the one hand, the system then lacks the singularity which makes it simpler on first sight. On the other hand we think that it becomes more difficult, because there will be inverse hyperbolic orbits, such that it is not sufficient to look for maxima of the Lagrangian in order to find periodic orbits and moreover a binary symbolic dynamics will not suffice.

## Acknowledgments

We would like to thank F. Steiner, P.H. Richter, A. Wittek, R. Aurich and T. Hesse for useful comments and discussions. We also thank the Deutsche Forschungsgemeinschaft for financial support.